\documentclass[arxiv]{w-art}
\usepackage{paper}
\newcommand{\supptitle}{%
    \begin{center}
        \Large Supplementary materials for ``A Bayesian framework for cost-effectiveness analysis with time-varying treatment decisions'' \\[0.35cm]
        \normalsize
        Esteban Fern\'andez-Morales\textsuperscript{1}, 
        Emily M. Ko\textsuperscript{2},
        Nandita Mitra\textsuperscript{3},
        Youjin Lee\textsuperscript{1}, and
        Arman Oganisian\textsuperscript{1,$\dagger$} \\[0.35cm]
        \normalsize
        \begin{tabular}{c}
            \textsuperscript{1}Department of Biostatistics, Brown University, USA \\
            \textsuperscript{2}Division of Gynecologic Oncology, University of Pennsylvania, Pennsylvania, USA \\
            \textsuperscript{3}Division of Biostatistics, University of Pennsylvania, Pennsylvania, USA \\
            \textsuperscript{$\dagger$}\textsf{e-mail: arman\_oganisian@brown.edu}
        \end{tabular}
        \vspace{5mm}
    \end{center}
}

\begin{document}

\DOIprefix{10.1002}
\DOIsuffix{arXiv.DOIsuffix}
\Volume{14}
\Issue{1}
\Year{2026}
\pagespan{1}{}


\title[Bayesian cost-effectiveness analysis with time-varying treatments]{A Bayesian framework for cost-effectiveness analysis with time-varying treatment decisions}

\author[Fern\'andez-Morales \textit{et al.}]{Esteban Fern\'andez-Morales\inst{1}}
\author[]{Emily M. Ko\inst{2}}
\author[]{Nandita Mitra\inst{3}}
\author[]{Youjin Lee\inst{1}}
\author[]{Arman Oganisian\inst{1,}\negthinspace\footnote{Corresponding author: \textsf{e-mail: arman\_oganisian@brown.edu}.}}

\address[\inst{1}]{Department of Biostatistics, Brown University, Rhode Island, USA}
\address[\inst{2}]{Division of Gynecologic Oncology, University of Pennsylvania, Pennsylvania, USA}
\address[\inst{3}]{Division of Biostatistics, University of Pennsylvania, Pennsylvania, USA}

\begin{abstract}
Cost-effectiveness analyses~(CEAs) compare the costs and health outcomes of treatment regimes to inform medical decisions.
With observational claims data, CEAs must address nonrandom treatment assignment, administrative censoring, and irregularly spaced medical visits that reflect the continuous timing of care and treatment initiation.
In high-risk, early-stage endometrial cancer~(HR-EC), adjuvant radiation is initiated at patient-specific times following hysterectomy, causing confounding between treatment and outcomes that can evolve with post-surgical recovery and clinical course.
Most existing CEA methods use point-treatment or discrete-time models.
However, point-treatment approaches break down with time-varying confounding, while discrete-time models bin continuous time, expand the data into a person-period format, and can induce zero-inflation by creating many intervals with no cost-accruing events.
We propose a Bayesian framework for CEAs with sequential decision-making that jointly models costs and event times in continuous time, accounts for administrative censoring, and supports dynamic treatment regimes with minimal parametric assumptions.
We use Bayesian g-computation to estimate \textit{causally interpretable} cost-effectiveness measures, including net monetary benefit, and to compare regimes through posterior contrasts.
We evaluate the finite-sample performance of the proposed method in simulations across censoring levels and compare it against discrete-time and fully parametric alternatives.
We then use SEER-Medicare data to assess the cost-effectiveness of initiating adjuvant radiation therapy within six months following hysterectomy among HR-EC patients.
\end{abstract}

\maketitle
\doublespacing

\section{Introduction}
\label{sec:introduction}

Cost-effectiveness analyses~(CEAs) compare the costs and health outcomes of one or more treatment regimes~\citep{Neumann2016CosteffectivenessHealth}.
They play a central role in drug pricing, coverage, and reimbursement decisions, shaping policy and resource allocation~\citep{Avancena2021ExaminingEquity,Clement2009UsingEffectiveness}.
A common summary measure in CEAs is the net monetary benefit~(NMB), which places costs and health outcomes on a common monetary scale through a willingness-to-pay~(WTP) threshold for an additional unit of health, such as a life-year~\citep{Varian1992MicroeconomicAnalysis,Nguyen2022GeneralFramework}.
An intervention is considered cost-effective if its expected health gains, valued at the WTP threshold, exceed its costs.

Despite their importance, CEAs present significant statistical challenges. 
Cost and survival outcomes are often correlated, costs are typically right-skewed, and survival outcomes are frequently subject to censoring~\citep{Handorf2019EstimatingCosteffectiveness,Li2016PropensityScore,Li2018DoublyRobust, Oganisian2020BayesianNonparametric}.
These challenges are exacerbated when analyses rely on observational data, such as insurance claims databases, where treatment assignment is nonrandom and confounding between treatment and outcomes is unavoidable.

This work is motivated by high-risk, early-stage endometrial cancer~(HR-EC). 
Standard treatment begins with total hysterectomy and bilateral salpingo-oophorectomy, followed by adjuvant radiation therapy~(aRT) in selected patients based on surgical staging and post-surgery recovery.
The \citeauthor{NationalComprehensiveCancerNetwork2023NCCNClinical} recommends two primary adjuvant options: external beam radiation therapy~(EBRT) and vaginal brachytherapy~(VBT).
In practice, the timing and choice of aRT are influenced by a patient's post-surgical recovery and evolving clinical status, resulting in potential time-varying confounding between treatment and outcomes.
We study the cost-effectiveness of initiating EBRT versus VBT within six months following hysterectomy using SEER-Medicare data, where administrative censoring is also common.

Standard point-treatment CEA methods~\citep{Oganisian2020BayesianNonparametric,Handorf2019EstimatingCosteffectiveness,Baio2014BayesianModels,Li2018DoublyRobust} are inadequate when treatment decisions and confounders change over time.
Such settings require causal methods that handle \textit{time-varying treatments and confounding} in either discrete or continuous time.
Discrete-time approaches often use g-estimation~\citep{Robins1986NewApproach} or Q-learning~\citep{Watkins1989LearningDelayed}.
Although g-estimation can adjust for time-varying confounding, most work targets either cumulative cost~\citep{Spieker2020NestedComputation, Spieker2018AnalyzingMedical} or discrete-time survival~\citep{Wen2021ParametricGformula,Chen2025FlexibleBayesian}.
Q-learning has been used to identify cost-effective regimes \citep{Illenberger2023IdentifyingOptimally}, but it targets \textit{policy optimization}.
Our goal is \textit{policy evaluation}: we compare prespecified regimes by estimating counterfactual cost and survival under each regime and then forming net-benefit contrasts.

Furthermore, discrete-time modeling also presents practical and statistical drawbacks.
Binning continuous time transforms the data into a person-period format, increasing computational and storage demands.
Coarse discretization can lead to information loss, while finer discretization often introduces zero inflation through intervals with no cost-accruing events, requiring additional modeling assumptions.
Crucially, survival time must be reconstructed from interval counts, which can introduce distortion when converted back to a continuous scale.

By contrast, continuous-time approaches model the timing between successive events directly---often referred to as \textit{gap times}---typically through transition- or hazard-based models~\citep{Xu2016BayesianNonparametric,Oganisian2024BayesianSemiparametric}.
This framework supports \textit{dynamic treatment regimes}~(DTRs), where decisions depend on a patient's evolving history~\citep{Murphy2003OptimalDynamic,Chakraborty2014DynamicTreatment}.
More importantly, survival can be expressed as the sum of event-specific gap times, allowing the full event history to inform inference without discretization.
This feature is critical in our setting, where medical encounters occur irregularly rather than at fixed intervals.
\citet{Hua2022PersonalizedDynamic} proposed a Bayesian approach for treatment optimization using a joint longitudinal and time-to-event model; however, their approach relies on parametric hazard specifications and targets optimal policies rather than counterfactual comparisons under regimes.

In this article, we develop a Bayesian joint modeling framework to estimate \textit{causally interpretable} cost-effectiveness measures for time-varying treatment decisions.
Our approach jointly models costs and gap times in continuous time, while relaxing parametric assumptions through semiparametric baseline hazard and mean-cost functions.
We use Bayesian g-computation to estimate counterfactual outcomes under DTRs, allowing us to quantify uncertainty via the posterior distribution.
This framework supports a broad class of estimands that are functions of total cost and survival time, including NMB, the estimation of which is often difficult using frequentist methods.
We target insurance claims settings with administrative censoring and time-varying confounding.

The remainder of the article is organized as follows.
Section~\ref{sec:notation} introduces the observed data structure and notation.
Section~\ref{sec:framework} presents the causal framework and defines the target estimands.
Section~\ref{sec:methods} describes the models for costs and gap times, while Section~\ref{sec:sampling} outlines the Bayesian g-computation algorithm used for posterior estimation.
Section~\ref{sec:simulation} evaluates the finite-sample performance of our proposed method through simulation.
Section~\ref{sec:application} applies the method to SEER-Medicare data to assess the cost-effectiveness of adjuvant EBRT versus VBT in HR-EC patients.
Section~\ref{sec:closing} concludes with a discussion and directions for future work.

\section{Observed Data Structure}
\label{sec:notation}

We observe a cohort of $n$ patients, indexed by $i = 1, \ldots, n$, through a sequence of healthcare encounters derived from insurance claims.
For patient $i$, let $V_{ij} \in \R^+$ denote the time of the $j$-th encounter, indexed by $j = 0, 1, \ldots, J_i$, with $V_{i0} = 0$ at hysterectomy.
At each encounter $j \geq 1$, we record covariates $\bsL_{ij} \in \cL$, a treatment-readiness indicator $Z_{ij} \in \qty{0, 1}$, a treatment decision $A_{ij} \in \cA$, and the cost for this encounter $Y_{ij} \in \R^+$.
The action space $\cA \subseteq \N$ is a small discrete set, and the covariate support $\cL \subseteq \R^P$ may include both continuous and discrete components, where $P$ is the number of covariates.
We denote baseline covariates measured at surgery by $\bsL_{i0}$ and treat them as a fixed subset of the $P$ covariates in $\bsL_{ij}$.
% We exclude surgery-related costs by setting $Y_{i0} = 0$.

In our SEER-Medicare application, $\cA = \qty{0, 1, 2}$ denotes no treatment, EBRT, and VBT, respectively.
Covariates $\bsL_{ij}$ include baseline characteristics such as age and cancer stage, along with time-varying measures such as comorbidity recorded at each encounter $j$.
We use $Z_{ij}$ to encode whether initiating aRT is clinically feasible at encounter $j$ (for example, adequate post-surgical recovery), so $A_{ij} \in \qty{0}$ if $Z_{ij} = 0$ and $A_{ij} \in \qty{1, 2}$ if $Z_{ij} = 1$.
In this application, aRT is initiated at most once and then remains fixed; that is, once $A_{ij} \in \qty{1, 2}$, it stays constant for all later encounters.
Our framework allows more general time-varying treatment rules, but this monotone-initiation structure is sufficient for the analysis.

Let $T_i \in \R^+$ denote the time to death and $C_i \in \R^+$ the censoring time.
The observed terminal time is $D_i = \min{T_i, C_i}$.
We define the $j$-th \textit{gap time} as $W_{ij} = \min{V_{ij}, D_i} - V_{i(j-1)}$, for $j = 1, \ldots, J_i$, which is the elapsed time between consecutive encounters, with the final gap possibly ending in death or censoring.
Let the transition indicator $\delta_{ij} \in \qty{0, 1, 2}$ denote the $j$-th event type: $\delta_{ij} = 0$ for censoring, $\delta_{ij} = 1$ for a new encounter, and $\delta_{ij} = 2$ for death.
For example, if $J_i = 2$ then $W_{i1} = V_{i1} - V_{i0}$ with $\delta_{i1} = 1$, and $W_{i2} = D_i - V_{i1}$ with $\delta_{i2} \in \qty{0, 2}$.

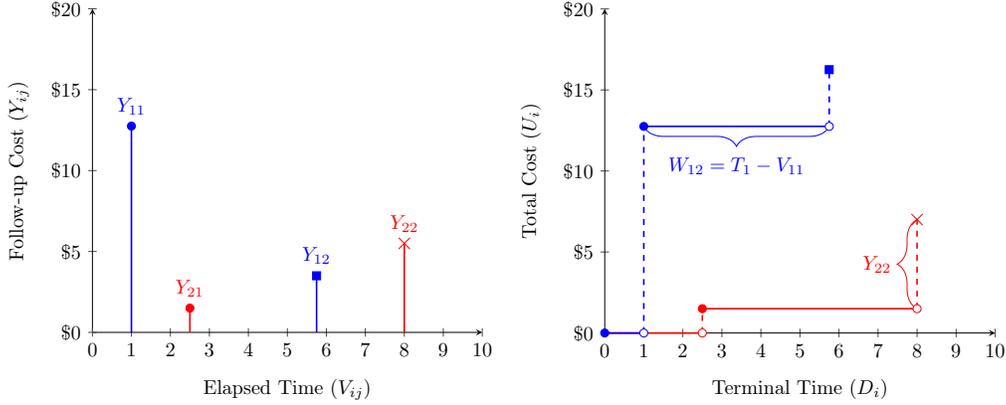
\begin{figure}[tbp]
    \centering
    \resizebox{0.45\linewidth}{!}{\begin{tikzpicture}
    \tikzmath{
        \v1 = 0;
        \v2 = 1;
        \t1 = 5.75;
        \x1 = \v2;
        \y1 = 12.75;
        \x2 = \t1;
        \y2 = 3.5;
        \s1 = 0;
        \s2 = 2.5;
        \c2 = 8;
        \w1 = \s2;
        \z1 = 1.5;
        \w2 = \c2;
        \z2 = 5.5;
    }
    \begin{axis}[
        axis x line=middle,
        axis y line=middle,
        ylabel={Follow-up Cost ($Y_{ij}$)},
        xlabel={Elapsed Time ($V_{ij}$)},
        xtick={0,1,...,10},
        extra x ticks={0},
        xlabel near ticks,
        ytick={0,5,...,20},
        extra y ticks={0},
        yticklabel={${\$\pgfmathprintnumber{\tick}}$},
        ylabel near ticks,
        xmax=10,
        ymax=20,
        xmin=0,
        ymin=0,
        clip=false
    ]
    \draw[thick,blue,solid] (\x1,0) -- (\x1,\y1);
    \draw[thick,blue,solid] (\x2,0) -- (\x2,\y2);
    \draw[thick,red,solid] (\w1,0) -- (\w1,\z1);
    \draw[thick,red,solid] (\w2,0) -- (\w2,\z2);
    \addplot[only marks,blue,mark=*,mark size=2pt] coordinates{(\x1, \y1)};
    \addplot[only marks,blue,mark=square*,mark size=2pt] coordinates{(\x2, \y2)};
    \addplot[only marks,red,mark=*,mark size=2pt] coordinates{(\w1, \z1)};
    \addplot[only marks,red,mark=x,mark size=4pt] coordinates{(\w2, \z2)};
    \node[fill=none,text=blue,scale=1,yshift=10pt] at (\x1,\y1) {$Y_{11}$};
    \node[fill=none,text=blue,scale=1,yshift=10pt] at (\x2,\y2) {$Y_{12}$};
    \node[fill=none,text=red,scale=1,yshift=10pt] at (\w1,\z1) {$Y_{21}$};
    \node[fill=none,text=red,scale=1,yshift=10pt] at (\w2,\z2) {$Y_{22}$};
    \end{axis}
\end{tikzpicture}}
    \resizebox{0.45\linewidth}{!}{\begin{tikzpicture}
    \tikzmath{
        \v1 = 0;
        \v2 = 1;
        \b1 = 12.75;
        \b2 = 3.5;
        \t1 = 5.75;
        \x1 = \v2;
        \y1 = \b1;
        \e1 = \v2 - \v1;
        \x2 = \t1;
        \y2 = \b1 + \b2;
        \e2 = \t1 - \v2;
        \s1 = 0;
        \s2 = 2.5;
        \d1 = 1.5;
        \d2 = 5.5;
        \c2 = 8;
        \w1 = \s2;
        \z1 = \d1;
        \w2 = \c2;
        \z2 = \d1 + \d2;
    }
    \begin{axis}[
        axis x line=middle,
        axis y line=middle,
        ylabel={Total Cost ($U_i$)},
        xlabel={Terminal Time ($D_i$)},
        xtick={0,1,...,10},
        extra x ticks={0},
        xlabel near ticks,
        ytick={0,5,...,20},
        extra y ticks={0},
        yticklabel={${\$\pgfmathprintnumber{\tick}}$},
        ylabel near ticks,
        xmax=10,
        ymax=20,
        xmin=0,
        ymin=0,
        clip=false
    ]
    \addplot[thick,red,domain=\s1:\s2] {0};
    \addplot[thick,red,domain=\s2:\c2] {\d1};
    \addplot[thick,blue,domain=\v1:\v2] {0};
    \addplot[thick,blue,domain=\v2:\t1] {\b1};
    \draw[thick,blue,dashed] (\x1,0) -- (\x1,\y1);
    \draw[thick,blue,dashed] (\x2,\y1) -- (\x2,\y2);
    \draw[thick,red,dashed] (\w1,0) -- (\w1,\z1);
    \draw[thick,red,dashed] (\w2,\z1) -- (\w2,\z2);
    \addplot[only marks,blue,mark=*,mark size=2pt] coordinates{(0, 0)};
    \addplot[only marks,blue,mark=*,mark size=2pt] coordinates{(\x1, \y1)};
    \addplot[only marks,blue,mark=square*,mark size=2pt] coordinates{(\x2, \y2)};
    \addplot[only marks,red,mark=*,mark size=2pt] coordinates{(\w1, \z1)};
    \addplot[only marks,red,mark=x,mark size=4pt] coordinates{(\w2, \z2)};
    \addplot[only marks,blue,fill=white,mark=*] coordinates{(\x1, 0)};
    \addplot[only marks,blue,fill=white,mark=*] coordinates{(\x2, \y1)};
    \addplot[only marks,red,fill=white,mark=*] coordinates{(\w1, 0)};
    \addplot[only marks,red,fill=white,mark=*] coordinates{(\w2, \z1)};
    \draw[blue,decorate,decoration={brace,mirror,amplitude=10pt}](\x1,\y1) -- (\x2,\y1) node[blue,midway,yshift=-20pt]{$W_{12} = T_1 - V_{11}$};
    \draw[red,decorate,decoration={brace,amplitude=10pt}](\w2,\z1) -- (\w2,\z2) node[red,midway,xshift=-20pt]{$Y_{22}$};
    \end{axis}
\end{tikzpicture}}
    \caption{Illustration of cost accrual in continuous time for two patients ($i = 1, 2$) with $J_i = 2$ post-surgery encounters. The left panel shows observed calendar times and associated costs, with nonterminal encounters indicated by circles~(\textcolor{blue}{$\bullet$}, \textcolor{red}{$\bullet$}).
    Patient~1 is observed at $V_{11} = 1$, while Patient~2 is observed at $V_{21} = 2.5$, with costs $Y_{11} = 12.75$ and $Y_{21} = 1.5$. Terminal events occur at death~(\textcolor{blue}{$\blacksquare$}) for Patient~1 at $T_1 = 5.75$ and censoring~(\textcolor{red}{$\times$}) for Patient~2 at $C_2 = 8$, with costs $Y_{12} = 3.5$ and $Y_{22} = 5.5$. The right panel shows cumulative cost accrual as a step function over time, where solid lines represent accumulated total cost and dashed lines indicate jumps at observed encounters. Gap times between events are $W_{11} = V_{11} - V_{10} = 1$, $W_{12} = T_1 - V_{11} = 4.75$, $W_{21} = V_{21} - V_{20} = 2.5$, and $W_{22} = C_2 - V_{21} = 5.5$.}
    \label{fig:cost-example}
\end{figure}

Total cost and observed follow-up time are sums of the gap-specific outcomes: $U_i = \sum_{j=1}^{J_i} Y_{ij}$ and $D_i = \sum_{j=1}^{J_i} W_{ij}$.
When death is observed, the right-hand side equals the true survival time (that is, $D_i = T_i$); otherwise, it equals the censoring time.
Figure~\ref{fig:cost-example} illustrates the cost accrual process and associated gap times.

The within-encounter temporal ordering is $(W_{ij}, \delta_{ij}) \to \bsL_{ij} \to Z_{ij} \to A_{ij} \to Y_{ij}$, with baseline covariates $\bsL_{i0}$ allowed to affect all subsequent variables.
We use overline and underline notation for segments of patient trajectories: $\obar{A}_{ij} = (A_{i1}, \ldots, A_{ij})$ and $\ubar{A}_{ij} = (A_{ij}, \ldots, A_{iJ_i})$, with analogous definitions for other variables.
Omitting the encounter index denotes the complete trajectory; for example, $\obar{A}_i = (A_{i1}, \ldots, A_{iJ_i})$.
We define the information available \textit{before} the treatment decision at encounter $j$ as $\bsH_{ij} = \qty(Z_{ij}, \bsL_{ij}, W_{ij}, \delta_{ij}, \bsS_{ij})$, where $\bsS_{ij}$ collects past observed data up to encounter~$j-1$, such that
\(
    \bsS_{ij} = \qty(\obar{Y}_{i(j-1)}, \obar{A}_{i(j-1)}, \obar{Z}_{i(j-1)}, \obar{\bsL}_{i(j-1)}, \obar{W}_{i(j-1)}, \obar{\delta}_{i(j-1)})
\)
for $j = 1, \ldots, J_i$.
Thus, $\bsH_{ij}$ contains the most recent event information $(W_{ij}, \delta_{ij})$ and current state $(Z_{ij}, \bsL_{ij})$, along with prior history.
The complete observed data for patient $i$ are 
\(
    \bsO_i = \qty(\obar{Y}_i, \obar{A}_i, \obar{Z}_i, \obar{\bsL}_i, \obar{W}_i, \obar{\delta}_i),
\)
and the full sample is $\bsO = (\bsO_1, \ldots, \bsO_n)$.

\section{Causal Estimands, Assumptions, and Identifiability}
\label{sec:framework}

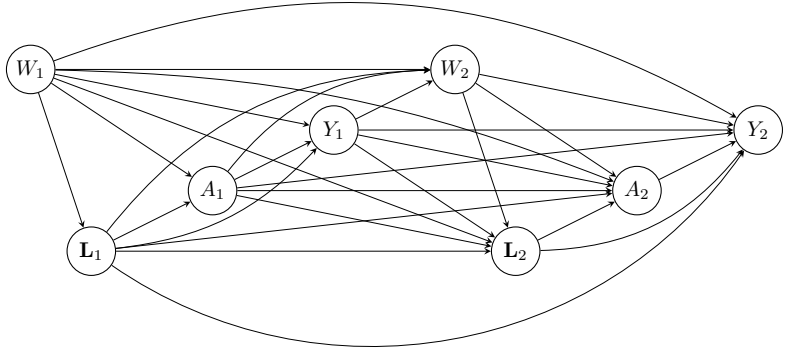
\begin{figure}[tbp]
    \centering
    \resizebox{0.7\linewidth}{!}{\begin{tikzpicture}[
    state/.style={draw, circle, minimum size=0.8cm, inner sep=1pt, font=\normalsize, fill=white, line width=0.5pt},
    >=stealth,
    node distance=1.5cm
]    
    % Visit j=1
    \node[state] (w1) at (-3, 2) {$W_1$};
    \node[state] (l1) at (-2, -1) {$\bsL_1$};
    \node[state] (a1) at (0, 0) {$A_1$};
    \node[state] (y1) at (2, 1) {$Y_1$};
    
    % Visit j=2
    \node[state] (w2) at (4, 2) {$W_2$};
    \node[state] (l2) at (5, -1) {$\bsL_2$};
    \node[state] (a2) at (7, 0) {$A_2$};
    \node[state] (y2) at (9, 1) {$Y_2$};
    
    % Paths from W1
    \path[->] (w1) edge (l1);
    \path[->] (w1) edge (a1);
    \path[->] (w1) edge (y1);
    \path[->] (w1) edge (w2);
    \path[->] (w1) edge (l2);
    \path[->] (w1) edge[bend left=10] (a2);
    \path[->] (w1) edge[bend left=25] (y2);
    
    % Paths from L1
    \path[->] (l1) edge (a1);
    \path[->] (l1) edge[bend right=20] (y1);
    \path[->] (l1) edge[bend left=25] (w2);
    \path[->] (l1) edge (l2);
    \path[->] (l1) edge (a2);
    \path[->] (l1) edge[bend right=45] (y2);
    
    % Paths from A1
    \path[->] (a1) edge (y1);
    \path[->] (a1) edge[bend left=25] (w2);
    \path[->] (a1) edge (l2);
    \path[->] (a1) edge (a2);
    \path[->] (a1) edge (y2);
    
    % Paths from Y1
    \path[->] (y1) edge (w2);
    \path[->] (y1) edge (l2);
    \path[->] (y1) edge (a2);
    \path[->] (y1) edge (y2);
    
    % Paths from W2
    \path[->] (w2) edge (l2);
    \path[->] (w2) edge (a2);
    \path[->] (w2) edge (y2);
    
    % Paths from L2
    \path[->] (l2) edge (a2);
    \path[->] (l2) edge[bend right=25] (y2);
    
    % Paths from A2
    \path[->] (a2) edge (y2);    
\end{tikzpicture}}
    \caption{Directed acyclic graph for $J = 2$ encounters illustrating time ordering and confounding, where baseline covariates $\bsL_0$ and treatment-readiness indicators $Z_j$ are omitted for clarity.}
    \label{fig:causal-dag}
\end{figure}

To draw causal conclusions from observational claims data, we formalize treatment strategies as DTRs and compare the outcomes they would generate.
We use the potential outcomes framework~\citep{Neyman1923ApplicationsTheorie,Rubin1974EstimatingCausal} to define causally interpretable cost-effectiveness estimands and state the assumptions under which they are identified from the observed data $\bsO$ (Section~\ref{sec:notation}).
When describing population-level quantities, we omit the patient index $i$.
Figure~\ref{fig:causal-dag} summarizes the causal structure.

Let $\obar{a}_j = (a_1, \ldots, a_j)$ denote a treatment history through encounter $j$.
The potential cost at encounter $j$ under history $\obar{a}_j$ is $Y_j(\obar{a}_j)$.
Because $(W_j, \delta_j)$ are realized before the treatment decision at encounter $j$, we index their potential outcomes by the preceding history $\obar{a}_{j-1}$:
$W_j(\obar{a}_{j-1})$ and $\delta_j(\obar{a}_{j-1})$, where $\delta_j(\obar{a}_{j-1}) = 1$ indicates a new encounter and $\delta_j(\obar{a}_{j-1}) = 2$ indicates death.
For presentation, we suppress censoring from the potential outcomes definitions here; censoring is handled through additional assumptions and the likelihood contributions in Supplementary Sections~\ref{supp:identification-assumptions} and~\ref{supp:likelihood-contributions}.
At the first post-surgery encounter ($j = 1$), no treatment decision has yet been made, so $W_1(\obar{a}_0) = W_1$ and $\delta_1(\obar{a}_0) = \delta_1$ are observed, with $\obar{a}_0 = 0$ by design.
Similarly, potential covariates and readiness indicators are $\bsL_j(\obar{a}_{j-1})$ and $Z_j(\obar{a}_{j-1})$.

Given a (possibly infinite) treatment sequence $\obar{a} = (a_1, a_2, \ldots)$, the potential number of encounters is $J(\obar{a}) = \max{j \in \N \colon \delta_j(\obar{a}_{j-1}) \neq 1}$.
The total potential cost and survival time under $\obar{a}$ are $U(\obar{a}) = \sum_{j=1}^{J(\obar{a})} Y_j(\obar{a}_j)$ and $T(\obar{a}) = \sum_{j=1}^{J(\obar{a})} W_j(\obar{a}_{j-1})$.

A DTR is a sequence of decision rules $\bsd = (d_1, d_2, \ldots)$, where each 
$d_j \colon \cH \to \cF(\bsH_j)$
maps the available history $\bsH_j$ (defined in Section~\ref{sec:notation})---with $\cH$ denoting the space of all possible histories---to a treatment decision in the \textit{feasible set} $\cF(\bsH_j) \subseteq \cA$.
The feasible set encodes restrictions implied by readiness and clinical practice; for example, $\cF(\bsH_j) = \qty{0}$ when $Z_j = 0$ and $\cF(\bsH_j) = \qty{1, 2}$ when $Z_j = 1$ in our application (see~\citet{Oganisian2024BayesianSemiparametric} for more examples).
Under $\bsd$, the induced treatment sequence is $A_j^{\bsd} \coloneqq d_j(\bsH_j)$, with $\obar{A}_j^{\bsd} \coloneqq (A_1^{\bsd}, \ldots, A_j^{\bsd})$.
This defines the shorthand potential outcomes $Y_j(\bsd) \equiv Y_j(\obar{a}_j^\bsd)$, $W_j(\bsd) \equiv W_j(\obar{a}_{j-1}^\bsd)$, $\delta_j(\bsd) \equiv \delta_j(\obar{a}_{j-1}^\bsd)$, and thus $J(\bsd) \equiv J(\obar{a}^{\bsd})$, $U(\bsd) \equiv U(\obar{a}^{\bsd})$, and $T(\bsd) \equiv T(\obar{a}^{\bsd})$.

In our SEER-Medicare application, we consider two illustrative regimes that initiate adjuvant therapy within six months post-hysterectomy once the patient is ready:
\begin{gather}
    \label{eq:regimes}
    d_1(\bsH_j) = 1 \cdot \one{Z_j = 1 \land V_j < 6}
    \quad \text{and} \quad
    d_2(\bsH_j) = 2 \cdot \one{Z_j = 1 \land V_j < 6},
\end{gather}
corresponding to initiating EBRT~($a_j = 1$) or VBT~($a_j = 2$), respectively, and assigning no aRT otherwise~($a_j = 0$).
Here, $V_j$ is time since hysterectomy, which can be determined by the observed history (that is, $V_j = \sum_{s=1}^j W_s$).

For a WTP threshold $\kappa \in \R^+$, the (counterfactual) monetary value~(MV) under regime $\bsd$ is
\begin{equation}
    \label{eq:mv}
    \psi_{\bsd}(\kappa) \coloneqq \E{\kappa T(\bsd) - U(\bsd)}.
\end{equation}
Here, $\psi_{\bsd}(\kappa)$ assigns financial value to survival at rate $\kappa$ and then subtracts the expected total cost.
Thus, $\psi_{\bsd}(\kappa) > 0$ indicates that, under $\bsd$, the survival benefit exceeds the expected cost at the WTP threshold $\kappa$.
To compare two regimes $\bsd$ and $\bsd'$, we use the NMB, 
\(
    \Psi_{\bsd, \bsd'}(\kappa) \coloneqq \psi_{\bsd}(\kappa) - \psi_{\bsd'}(\kappa).
\)
A positive $\Psi_{\bsd, \bsd'}(\kappa)$ favors $\bsd$ over $\bsd'$ at WTP $\kappa$, while a negative value favors $\bsd'$.

More generally, our framework targets estimands of the form
\(
    \psi_{\bsd}(g) \coloneqq \E{g\qty(\obar{Y}(\bsd), \obar{W}(\bsd), J(\bsd))},
\)
for any measurable function $g$ (including time-restricted versions under a finite horizon $\tau$).
Identification of $\psi_{\bsd}(\kappa)$ or $\psi_{\bsd}(g)$ requires the joint distribution of the counterfactual trajectory 
$\qty(\obar{Y}(\bsd), \obar{W}(\bsd), \obar{\delta}(\bsd))$, 
which is not identifiable from the observed data without additional assumptions.
Under the convention that censoring is handled separately, $J(\bsd)$ and $\obar{\delta}(\bsd)$ determine one another:
$J(\bsd) = r$ implies $\obar{\delta}(\bsd) = (\obar{1}_{r-1}, 2)$, and conversely.
We denote this terminal transition pattern by $\bse_r \coloneqq (\obar{1}_{r-1}, 2)$, representing $r - 1$ transitions to a new encounter followed by death.

We adopt the following standard assumptions for all $j \in \N$:
\begin{enumerate}
    \item \label{asm:treatment-positivity}
    \textbf{Treatment positivity:} 
    $P(A_j = a_j \mid \bsH_j = \bsh_j) > 0$ for all $a_j \in \cF(\bsh_j)$ and $\bsh_j \in \cH$.
    \item \label{asm:sequential-ignorability}
    \textbf{Ignorability:} 
    $\ubar{Y}_j(\obar{a}_j), \ubar{W}_{j+1}(\obar{a}_j), \ubar{\delta}_{j+1}(\obar{a}_j), \ubar{\bsL}_{j+1}(\obar{a}_j), \ubar{Z}_{j+1}(\obar{a}_j) \indep A_j \mid \bsH_j$ for all $a_j \in \cF(\bsH_j)$.
\end{enumerate}
Together with the stable unit treatment value assumption (no interference and well-defined treatments), Assumptions~\ref{asm:treatment-positivity} and~\ref{asm:sequential-ignorability} link counterfactual and observed quantities~\citep{Rubin1980RandomizationAnalysis}.
We also assume non-informative censoring and censoring positivity conditional on the observed history, with a formal statement given in Supplementary Section~\ref{supp:identification-assumptions}.
Collectively, these assumptions ensure confounding is addressed by the observed longitudinal history, rule out systematic differences between censored and uncensored patients, and guarantee sufficient overlap to identify the counterfactual quantities of interest and estimate them from the observed data.

Under these conditions, the joint density of the counterfactual outcomes for a given number of encounters $J(\bsd) = r$ can be written as the following (continuous-time, joint-outcome) g-formula:
\begin{equation}
    \label{eq:g-formula}
    \begin{split}
        & f(\obar{y}, \obar{w}, \bse_r) \\
        & \quad = \int_{\cL_0} \int_{\obar{\cX}_r} \Bigg[ f(y_r \mid A_r = a_r, \bsH_r = \bsh_r) f(\bsx_r \mid W_r = w_r, \delta_r = 2, \bsS_r = \bss_r) f(w_r, \delta_r = 2 \mid \bsS_r = \bss_r) \\
        & \qquad\qquad \times \prod_{j=1}^{r-1} f(y_j \mid A_j = a_j, \bsH_j = \bsh_j) \\
        & \qquad\qquad \times \prod_{j=1}^{r-1} f(\bsx_j \mid W_j = w_j, \delta_j = 1, \bsS_j = \bss_j) \\ 
        & \qquad\qquad \times \prod_{j=1}^{r-1} f(w_j, \delta_j = 1 \mid \bsS_j = \bss_j) \Bigg] f(\bsl_0) \, \dd \obar{\bsx}_r \, \dd \bsl_0,
    \end{split}
\end{equation}
where $\cL_0$ is the support of $\bsL_0$, $\cX_j \coloneqq \cL \times \qty{0, 1}$ is the support of $\bsX_j = (\bsL_j, Z_j)$, and $a_j = d_j(\bsh_j)$ under regime $\bsd$.
Equation~\eqref{eq:g-formula} expresses the joint counterfactual distribution in terms of observed-data conditionals, allowing identification of $\psi_{\bsd}(\kappa)$ and $\psi_{\bsd}(g)$ using $\bsO$.
In Sections~\ref{sec:methods} and~\ref{sec:sampling}, we specify Bayesian models for the conditional densities and approximate the resulting functional using Bayesian g-computation.
A derivation of~\eqref{eq:g-formula} is provided in Supplementary Section~\ref{supp:g-formula}.

\section{Bayesian Joint Modeling of Gap Times and Costs}
\label{sec:methods}

We specify Bayesian models for each component of the g-formula in~\eqref{eq:g-formula}.
For computational tractability, we adopt two \textit{modeling restrictions}---time homogeneity and a first-order Markov property---which reduce the dimension of the conditioning sets in~\eqref{eq:g-formula}.
% Time homogeneity assumes that model parameters do not vary with the encounter index, and the first-order Markov restriction assumes that, given the most recent state, the distribution of future variables does not depend on earlier history.
These restrictions are application-specific; they yield parsimonious models and stabilize estimation when trajectories are long and the conditioning sets become high-dimensional.
Both can be relaxed (for example, time-varying coefficients or higher-order lags) at the cost of additional parameters and increased computational burden.
Accordingly, for $j = 1, \ldots, J$, we define the one-lag state and the corresponding history as
$\bsS_j = (Y_{j-1}, A_{j-1}, Z_{j-1}, \bsL_{j-1}, W_{j-1}, \delta_{j-1})$
and
$\bsH_j = \qty(Z_j, \bsL_j, W_j, \delta_j, \bsS_j)$,
respectively.

We present the joint model for the gap time and cost processes below.
We also require models for the covariate process $\bsL_j$ and treatment readiness $Z_j$ to implement g-computation.
% Because these components are nuisance models whose specification depends on the chosen covariates and data structure, we defer details to the simulation study (Section~\ref{sec:simulation}), and note that the application uses the same modeling template.
Because these components are nuisance models whose specification depends on the chosen covariates and data structure, we defer details to Supplementary Section~\ref{supp:covariates-treatment-readiness}.

We model the distribution of gap times using a cause-specific proportional hazards model indexed by event type $k \in \qty{1, 2}$ and the distribution of costs using a proportional means model:
\begin{equation}
    \begin{split} 
        \label{eq:joint-model}
        h(w_j, \delta_j = k \mid \bsS_j = \bss_j)
            &= h_{0k}(w_j) \exp(\bss_j' \bsvarphi_k), \\
        m(w_j \mid \delta_j = k, A_j = a_j, \bsH_j = \bsh_j)
            &= m_0(w_j) \exp(\beta_k + \bsa_j' \bsbeta_A + \bsh_j' \bsbeta_H).
    \end{split}
\end{equation}
Here, $h(\cdot)$ denotes the cause-specific hazard for the gap time $W_j$, while $m(\cdot)$ denotes the conditional \textit{mean of the cost outcome} $Y_j$, that is,
\begin{equation*}
    \mu_j 
        \coloneqq m(w_j \mid \delta_j = k, A_j = a_j, \bsH_j = \bsh_j)
        \equiv \E{Y_j \mid W_j = w_j, \delta_j = k, A_j = a_j, \bsH_j = \bsh_j}.
\end{equation*}
The regression coefficients $\bsvarphi_k$ capture the effects of the previous state $\bsS_j$ on the hazard of event type~$k$, while $\bsbeta = (\beta_k, \bsbeta_A, \bsbeta_H)$ capture the effects of event type $k$, treatment $A_j$, and observed history $\bsH_j$ on the mean cost, respectively.
We represent the treatment at encounter $j$ either as a scalar $a_j$ or as a one-hot vector $\bsa_j$ over the action space $\cA$; for example, $\bsa_j = (0, 1, 0)$ for EBRT or $\bsa_j = (0, 0, 1)$ for VBT.
% Furthermore, the proportional means model includes an event-type shift $\beta_k$, since empirical results (Supplementary Figure~\ref{fig:event-specific-cost-density}) show systematic differences between costs at non-terminal encounters and those recorded at death.

The functions $m_0(\cdot)$ and $h_{0k}(\cdot)$ denote the baseline mean-cost modifier and the baseline cause-specific hazard, respectively.
We model both using piecewise constant functions: 
\begin{equation*}
    m_0(w_j) = \sum_{q=1}^Q m_{0q} \one{w_j \in \cI_q}
    \quad \text{and} \quad
    h_{0k}(w_j) = \sum_{q=1}^Q h_{0kq} \one{w_j \in \cI_q},
\end{equation*}
where $(\cI_1, \ldots, \cI_Q)$ partitions the follow-up time into $Q$ evenly spaced, non-overlapping intervals~\citep{Oganisian2024BayesianSemiparametric}.
The vectors $\bsm_0 = (m_{01}, \ldots, m_{0Q})$ and $\bsh_{0k} = (h_{0k1}, \ldots, h_{0kQ})$ denote the interval-specific baseline mean-cost modifiers and cause-specific hazard rates, respectively.
For consistency, we use the same partition in both models.

This piecewise formulation is useful when the dependence of mean cost on gap time is unknown. 
For example, mean costs may increase with longer gaps (more complex clinical visits) or decrease with longer gaps (fewer acute events), or show little dependence at all.
In settings where this dependence is negligible, we may simplify by setting $m_0(w_j) \equiv 1$, reducing~\eqref{eq:joint-model} to a standard regression model for the mean cost.
Similarly, the piecewise hazard avoids strong parametric assumptions (for example, exponential or Weibull), allowing $h_{0k}$ to flexibly approximate the true hazard as $Q$ increases.

The gap time density follows the standard cause-specific factorization.
Suppressing conditioning terms for brevity, the cause-specific density is
$f(w_j, \delta_j = k \mid -) = h(w_j, \delta_j = k \mid -) S(w_j \mid -)$,
where the overall survival function is
$S(w_j \mid -) = \exp(-\int_0^{w_j} \sum_{k \in \qty{1, 2}} h(u, \delta_j = k \mid -) \, \dd u)$.
We model $Y_j$ using a gamma distribution with mean-variance parameterization, $Y_j \sim \GammaDist(\mu_j^2 / \zeta, \mu_j / \zeta)$, so that $\E{Y_j} = \mu_j$ and $\Var{Y_j} = \zeta$.
This provides a positive-support likelihood for right-skewed claims costs; other likelihoods (for example, log-normal or inverse gamma) can be substituted.
Our use of the term ``semiparametric'' refers to the flexible baseline functions in the joint model---the cause-specific hazards $h_{0k}(\cdot)$ and the baseline mean-cost modifier $m_0(\cdot)$---which we approximate with piecewise constant functions (using smoothing priors outlined in Supplementary Section~\ref{supp:smoothing-priors}) to reduce sensitivity to misspecified functional forms.

\section{Posterior Computation of Causal Estimands}
\label{sec:sampling}

We estimate the posterior distribution of the causal estimand $\psi_{\bsd}(g)$ introduced in Section~\ref{sec:framework}.
Let $\bstheta$ denote the collection of parameters indexing the models in Section~\ref{sec:methods}, including the gap time, cost, covariate, and readiness components; we write 
$\bstheta = (\bsh_{00}, \bsvarphi_0, \bsh_{01}, \bsvarphi_1, \bseta, \bsphi, \bsm_0, \bsbeta, \zeta)$,
where $(\bseta, \bsphi)$ parameterize the covariate and readiness models (Supplementary Section~\ref{supp:covariates-treatment-readiness}).
We provide the likelihood, priors, and posterior specification in Supplementary Sections~\ref{supp:likelihood-contributions}--\ref{supp:smoothing-priors}.

Because the posterior is not available in closed form, we use \texttt{Stan}~\citep{Carpenter2017StanProbabilistic} to draw $\bstheta^{(1)}, \ldots, \bstheta^{(M)}$ via Hamiltonian Monte Carlo~\citep{Homan2014NoUturnSampler}.
For each posterior draw $\bstheta^{(m)}$ and regime $\bsd$, we approximate the g-formula in~\eqref{eq:g-formula} by Monte Carlo g-computation.
Algorithm~\ref{alg:g-computation} summarizes the Bayesian g-computation procedure used to generate posterior draws of $\psi_{\bsd}(g)$.
Specifically, we simulate $B$ independent counterfactual trajectories under $\bsd$ and $\bstheta^{(m)}$, yielding draws $\qty{U^{(b)}, T^{(b)}}_{b=1}^B$, and compute 
\begin{equation*}
    \psi_{\bsd}^{(m)}(g)
        \approx \frac{1}{B} \sum_{b=1}^B g\qty(U^{(b)}, T^{(b)}).
\end{equation*}
Repeating this procedure for $m = 1, \ldots, M$ yields posterior draws $\psi_{\bsd}^{(1)}(g), \ldots, \psi_{\bsd}^{(M)}(g)$.
For MV, we take $g_\kappa(u, t) = \kappa t - u$, so that 
$\psi_{\bsd}^{(m)}(\kappa) \equiv \psi_{\bsd}^{(m)}(g_\kappa)$.
\begin{algorithm}[tbp]
    \caption{Bayesian g-computation for approximating causal estimands.}
    \label{alg:g-computation}
    \begin{algorithmic}[1]
        \State \textbf{Input:} Posterior draw $\bstheta^{(m)}$, regime $\bsd$, horizon $\tau$ (optional), Monte Carlo size $B$
        \State \textbf{Output:} Posterior draw $\psi_{\bsd}^{(m)}(g)$
        \For{$b = 1$ to $B$}
            \State Draw baseline covariates: $\bsL_0 \sim f(\bsl_0 \mid \bseta^{(m)})$
            \State Initialize: $j \gets 1$, $V_0 \gets 0$, $\delta_1 \gets 1$, $\bsS_1 \gets \bsL_0$ 
            \While{$\delta_j = 1$} \Comment{iterate until a terminal event (death or administrative horizon)}
                \State Draw gap time to next encounter: 
                $W_{Y_j} \sim f(w_j, \delta_j = 1 \mid \bsS_j, \bsh_{00}^{(m)}, \bsvarphi_0^{(m)})$
                \State Draw gap time to death: 
                $W_{T_j} \sim f(w_j, \delta_j = 2 \mid \bsS_j, \bsh_{01}^{(m)}, \bsvarphi_1^{(m)})$
                \State Set $W_j^\star \gets \min{W_{T_j}, W_{Y_j}}$ and $\delta_j^\star \gets \one{W_{Y_j} < W_{T_j}} + 2 \cdot \one{W_{T_j} \geq W_{Y_j}}$
                \If{a horizon $\tau$ is imposed \textbf{and} $V_{j-1} + W_j^\star > \tau$}
                    \State Set $W_j \gets \tau - V_{j-1}$ and $\delta_j \gets 0$
                    \Comment{truncate follow-up at $\tau$ and stop the trajectory}
                \Else
                    \State Set $W_j \gets W_j^\star$ and $\delta_j \gets \delta_j^\star$
                \EndIf
                \State Update calendar time: $V_j \gets V_{j-1} + W_j$
                \State Draw covariates: $\bsL_j \sim f(\bsl_j \mid W_j, \delta_j, \bsS_j = \bss_j, \bseta^{(m)})$
                \State Draw readiness: $Z_j \sim f(z_j \mid \bsL_j, W_j, \delta_j, \bsS_j, \bsphi^{(m)})$
                \State Update history: $\bsH_j \gets \qty(Z_j, \bsL_j, W_j, \delta_j, \bsS_j)$ 
                \State Assign treatment: $A_j^{\bsd} \gets d_j(\bsH_j)$    
                \State Compute mean cost: 
                $\mu_j \gets m(W_j \mid \delta_j, A_j^{\bsd}, \bsH_j, \bsm_0^{(m)}, \bsbeta^{(m)})$
                \State Draw cost: $Y_j \sim \GammaDist(\mu_j^2 / \zeta^{(m)}, \mu_j / \zeta^{(m)})$
                \State Update state: $\bsS_{j+1} \gets (Y_j, A_j^{\bsd}, Z_j, \bsL_j, W_j, \delta_j)$
                \State Increment: $j \gets j + 1$
            \EndWhile
            \State Compute totals: $U^{(b)} \gets \sum_{s=1}^{j} Y_s$, $T^{(b)} \gets \sum_{s=1}^{j} W_s$
        \EndFor
        \State Compute estimand draw: 
        $\psi_{\bsd}^{(m)}(g) \gets \frac{1}{B} \sum_{b=1}^B g\qty(U^{(b)}, T^{(b)})$
    \end{algorithmic}
\end{algorithm}

To compare two regimes $\bsd$ and $\bsd'$, we compute both estimands under the \textit{same} posterior draw $\bstheta^{(m)}$.
We then form posterior draws of the contrast as 
$\Psi_{\bsd, \bsd'}^{(m)}(g) = \psi_{\bsd}^{(m)}(g) - \psi_{\bsd'}^{(m)}(g)$ 
for~$m = 1, \ldots, M$.
We summarize posterior uncertainty using standard summaries such as posterior means and 95\% credible intervals.

\section{Simulation Study}
\label{sec:simulation}

We conduct a simulation study to evaluate the finite-sample performance of the proposed framework.
Each experiment consists of 1,000 independently replicated datasets with $n = 1,000$ patients.
We consider two censoring rates: 10\% (low) and 50\% (high).
The data-generating process is designed to mimic our SEER-Medicare application by simulating encounter-level claims indexed by $j \in \N$; full details are provided in Supplementary Section~\ref{supp:simulation-details}.

For each patient, we generate three baseline covariates (one binary and two continuous) and two time-varying covariates (one binary and one continuous), with the time-varying covariates depending on past history and treatment.
Treatment initiation is deterministic after surgery ($Z_j = 1$ for all $j \geq 1$), reflecting that aRT is not delivered at hysterectomy.
Treatment assignment is Bernoulli, with probability depending on current covariates $\bsL_j$ and lagged history $(A_{j-1}, \bsL_{j-1})$.
Once initiated ($A_j = 1$), treatment remains fixed until censoring or death.
Costs and gap times are generated from the proportional means and cause-specific proportional hazards models in~\eqref{eq:joint-model}, with a Gompertz baseline mean-cost function and Weibull baseline hazards.
Censoring times are sampled independently from an exponential distribution, with the rate calibrated to achieve the target censoring proportion.
For simplicity, we impose time homogeneity and a first-order Markov structure in the data-generating process, with $\bsH_j = (\bsL_j, \bsS_j)$ and $\bsS_j = (A_{j-1}, \bsL_{j-1})$.

Our target estimand is the $\tau$-restricted expected NMB, $\Psi_{d_1, d_0}(\kappa)$, comparing an always-treated regime $d_1(\bsh_j) \equiv 1$ to a never-treated regime $d_0(\bsh_j) \equiv 0$, with MV defined in~\eqref{eq:mv}.
We set $\kappa = 1$ and restrict follow-up to $\tau = 3$.
Setting $\kappa = 1$ provides a natural benchmark in which one unit of survival is weighted equally to one unit of cost on the analysis scale.
The true value of $\Psi_{d_1, d_0}(\kappa)$ is approximated using a Monte Carlo benchmark based on simulating outcomes for 1,000,000 patients under each regime.

We compare five estimation strategies. 
These include (i) correctly specified Bayesian parametric models; (ii) correctly specified maximum likelihood (ML) models; (iii) the proposed Bayesian joint model with piecewise baseline hazards and mean-cost functions; (iv) misspecified Bayesian parametric models assuming a Weibull mean-cost function and exponential hazards; and (v) discrete-time ML models implemented via modified g-computation~\citep{Spieker2018AnalyzingMedical,Spieker2020NestedComputation}.
Piecewise baseline functions use $Q = 10$ intervals, and the discrete-time comparator partitions follow-up into 25 intervals.
Performance is summarized using relative bias, standard error~(SE), root mean square error~(RMSE), interval width, and 95\% coverage rate~(CR).
Bayesian inference uses one Hamiltonian Monte Carlo chain with $M = 2,000$ post-warmup draws after 1,000 warmup iterations; frequentist uncertainty is obtained from 2,000 bootstrap resamples.
For each posterior (or bootstrap) draw, g-computation uses $B = 20,000$ simulated trajectories.

Table~\ref{tab:simulation-results} reports results across censoring rates and estimation strategies.
Relative to the correctly specified models, the proposed Bayesian joint model achieves comparable bias, standard errors, RMSE, interval widths, and coverage under both low and high censoring, with differences consistent with Monte Carlo variability.
As expected, heavier censoring modestly increases bias and widens intervals due to reduced information.
When baseline hazards and mean-cost functions are misspecified, bias increases and interval widths grow, producing coverage above 95\%, consistent with inflated uncertainty.
The discrete-time comparator performs the worst, with substantial bias, wide intervals, and low coverage.
This sensitivity reflects the need for fine discretization to approximate continuous-time dynamics and the modeling burden introduced by person-period expansion.
In contrast, the proposed approach performs well without requiring discretization or ad hoc adjustments for sparsely observed intervals.
\begin{table}[tbp]
    \centering
    \caption{Finite-sample performance across censoring rates (1,000 replications; $n = 1,000$).}
    \label{tab:simulation-results}
    \begin{tabularx}{\linewidth}{Lrrrrr}
        \toprule
        \textbf{Method} & \textbf{\% Bias} & \textbf{SE} & \textbf{RMSE} & \textbf{Width} & \textbf{CR (\%)} \\ 
        \midrule
        \multicolumn{6}{l}{\textit{Low censoring (10\%).}} \\ 
        Proposed Bayesian joint model             & -0.01 & 0.080 & 0.106 & 0.314 & 96.28 \\
        Bayesian parametric (correctly-specified) & -0.33 & 0.079 & 0.105 & 0.309 & 96.50 \\
        ML parametric (correctly-specified)       & -0.75 & 0.079 & 0.106 & 0.308 & 96.00 \\
        Misspecified Bayesian parametric          & 2.76  & 0.088 & 0.112 & 0.346 & 98.00 \\
        Discrete-time ML                          & -39.57 & 0.123 & 0.244 & 0.483 & 64.30 \\
        \addlinespace
        \multicolumn{6}{l}{\textit{High censoring (50\%).}} \\
        Proposed Bayesian joint model             & -0.50 & 0.099 & 0.131 & 0.390 & 95.80 \\
        Bayesian parametric (correctly-specified) & -1.93 & 0.098 & 0.129 & 0.384 & 95.60 \\
        ML parametric (correctly-specified)       & -2.57 & 0.097 & 0.130 & 0.381 & 95.60 \\
        Misspecified Bayesian parametric          & 4.87  & 0.109 & 0.137 & 0.428 & 98.20 \\
        Discrete-time ML                          & -78.07 & 0.193 & 0.442 & 0.759 & 47.10 \\
        \bottomrule
    \end{tabularx}
    \note[Notes:]{Entries are averages over 1,000 replications. The estimand is the $\tau$-restricted NMB contrast with $\tau = 3$ and $\kappa = 1$; the Monte Carlo benchmark is $\Psi_{d_1, d_0}(1) \approx -0.494$. Percent bias is relative to the benchmark. SE is the mean estimated standard error across replications. RMSE is computed as $\sqrt{\mathrm{SE}^2 + \mathrm{Bias}^2}$. Width and CR correspond to 95\% credible or bootstrap confidence intervals.}
\end{table}

\section{Application to High-Risk, Early-Stage Endometrial Cancer}
\label{sec:application}

We apply the proposed framework to evaluate the cost-effectiveness of post-hysterectomy aRT in HR-EC.
Our analysis uses SEER-Medicare-linked data on 19,399 patients who underwent primary hysterectomy for endometrial cancer between 2000 and 2017 and were followed through inpatient and outpatient Medicare claims, with observation through December 19, 2019.
To target the HR-EC population, we restrict to stage IA/IB patients with grade 3 disease, yielding a final sample of $n = 2,027$.
We treat claims as proxies for healthcare encounters and use them to define encounter times.

The initiation and choice of aRT depend on patient characteristics and clinical course, creating the potential for confounding.
We adjust for seven baseline covariates measured at surgery (five demographic and two cancer-related variables) and two time-varying covariates updated at each encounter: (i)~the Charlson Comorbidity Index, dichotomized as low~($< 3$) versus high~($\geq 3$), and (ii)~an indicator of additional therapies (chemotherapy or other radiation therapies).
Table~\ref{tab:patient-characteristics} reports baseline and aggregate characteristics by treatment group.
Standardized mean differences indicate imbalance in most covariates (except ethnicity and marital status), motivating adjustment.
Among the 2,027 patients, 724~(35.72\%) received EBRT and 425~(20.97\%) received VBT.
Administrative censoring occurred in 1,076 patients~(53.08\%), consistent with the high-censoring scenario in Section~\ref{sec:simulation}.
\begin{table}[tbp]
    \centering
    \caption{Patient characteristics by aRT group.}
    \label{tab:patient-characteristics}
    \begin{adjustbox}{width=\linewidth, center}
        \begin{tabular}{l r r r r r}
            \toprule
            \textbf{Characteristic} & \textbf{Overall} & \textbf{None} & \textbf{EBRT} & \textbf{VBT} & \textbf{SMD} \\
            \cmidrule(lr){2-5}
             & $n=2,027$ & $n=878$ & $n=724$ & $n=425$ & \\
            \midrule
            \multicolumn{6}{l}{\textit{Demographics and baseline clinical factors}} \\
            Age (years) & 73.8 (8.3) & 74.8 (9.1) & 73.1 (7.6) & 73.2 (7.6) & 0.137 \\
            White & 1,744 (86.0\%) & 750 (85.4\%) & 617 (85.2\%) & 377 (88.7\%) & 0.069 \\
            Married & 859 (42.4\%) & 378 (43.1\%) & 294 (40.6\%) & 187 (44.0\%) & 0.046 \\
            High socioeconomic status & 990 (48.8\%) & 402 (45.8\%) & 333 (46.0\%) & 255 (60.0\%) & 0.192 \\
            Metropolitan residence & 1,686 (83.2\%) & 716 (81.5\%) & 597 (82.5\%) & 373 (87.8\%) & 0.115 \\
            Nodes examined & 1,580 (77.9\%) & 642 (73.1\%) & 549 (75.8\%) & 389 (91.5\%) & 0.331 \\
            Stage IA & 1,233 (60.8\%) & 634 (72.2\%) & 322 (44.5\%) & 277 (65.2\%) & 0.388 \\
            \addlinespace
            \multicolumn{6}{l}{\textit{Outcomes and other variables}} \\
            Censored (dropout) & 1,076 (53.1\%) & 485 (55.2\%) & 304 (42.0\%) & 287 (67.5\%) & 0.351 \\
            Survival time (years) & 7.18 (4.77) & 7.65 (4.94) & 6.99 (4.80) & 6.56 (4.25) & 0.156 \\
            Total cost (1,000 USD) & 47.64 (67.59) & 42.78 (64.14) & 56.19 (75.03) & 43.14 (59.38) & 0.130 \\
            % Claim count & 6.48 (5.09) & 5.81 (4.74) & 7.58 (5.81) & 5.99 (4.09) & 0.231 \\
            Gap time (years) & 1.52 (1.39) & 1.86 (1.68) & 1.20 (1.07) & 1.35 (0.99) & 0.323 \\
            Cost per claim (1,000 USD) & 5.51 (5.07) & 5.30 (5.60) & 5.85 (4.36) & 5.38 (5.02) & 0.076 \\
            High comorbidity & 970 (47.9\%) & 449 (51.1\%) & 346 (47.8\%) & 175 (41.2\%) & 0.134 \\
            Other therapies received & 73 (3.6\%) & 16 (1.8\%) & 43 (5.9\%) & 14 (3.3\%) & 0.145 \\
            Initiated aRT & 1,149 (56.7\%) & -- & -- & -- & -- \\
            \bottomrule
        \end{tabular}
    \end{adjustbox}
    \note[Notes:]{Continuous variables are mean (SD); categorical variables are $n$ (\%). SMD denotes the standardized mean difference comparing each treated group to no aRT; values $> 0.1$ commonly indicate imbalance. Abbreviations: SD, standard deviation; EBRT, external beam radiation therapy; VBT, vaginal brachytherapy; aRT, adjuvant radiation therapy.}
\end{table}

We compare two treatment strategies in~\eqref{eq:regimes} that initiate EBRT ($d_1$) or VBT ($d_2$) within six months of hysterectomy.
For a WTP threshold $\kappa$, the three-year MV under strategy $\bsd$ is $\psi_{\bsd}(\kappa)$, and the NMB for EBRT versus VBT is 
$\Psi_{d_1, d_2}(\kappa) = \psi_{d_1}(\kappa) - \psi_{d_2}(\kappa)$.
Positivity is supported by observed initiation patterns: 588 (81.2\%) EBRT patients and 404 (95.1\%) VBT patients initiated within six months (Figure~\ref{fig:treatment-positivity}).

We fit the Bayesian joint model in~\eqref{eq:joint-model}, using piecewise constant baseline functions for the cause-specific hazards and the baseline mean-cost modifier.
Models for the covariate process and readiness indicator follow the same template as in Section~\ref{sec:simulation}, with further details in Supplementary Section~\ref{supp:application-details}.
Posterior inference uses four Hamiltonian Monte Carlo chains, each with 1,000 warmup iterations followed by $M = 1,000$ retained draws.
For each posterior draw, we approximate $\psi_{d_1}(\kappa)$ and $\psi_{d_2}(\kappa)$ via g-computation using $B = 20,000$ simulated trajectories under a three-year horizon, and obtain samples of $\Psi_{d_1, d_2}(\kappa)$ by differencing within each draw.

\begin{figure}[tbp]
    \centering
    \includegraphics[width=0.8\linewidth]{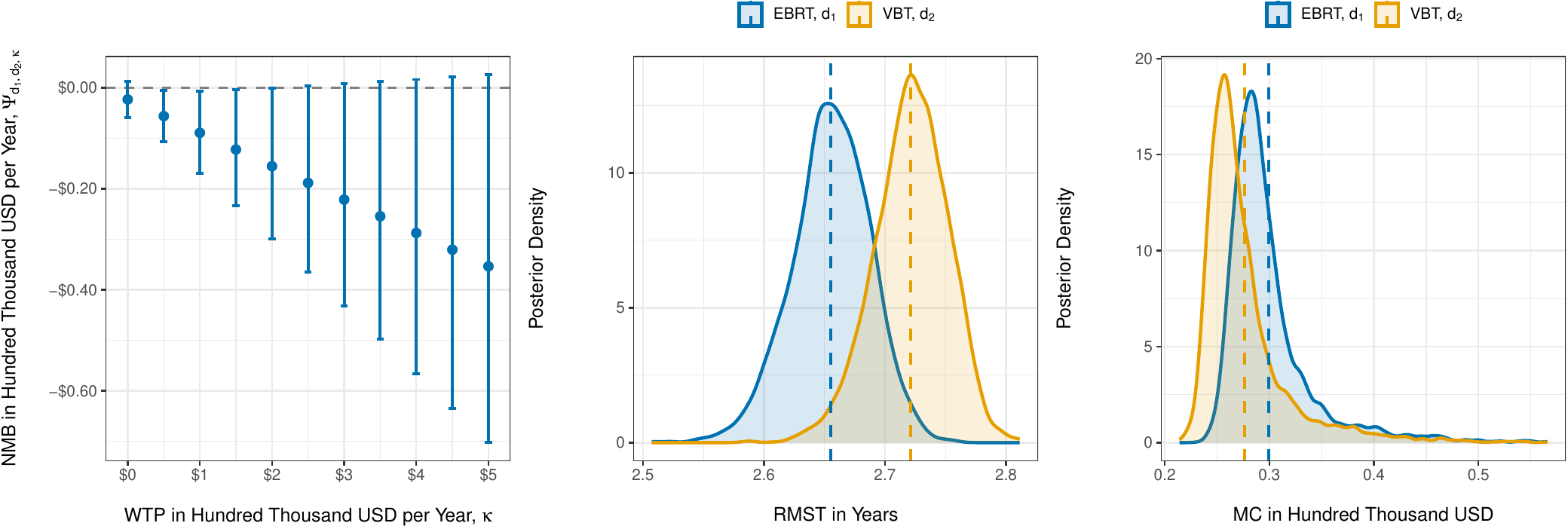}
    \caption{Posterior summaries for EBRT versus VBT. Left: posterior mean and 95\% credible interval for the NMB $\widehat{\Psi}_{d_1, d_2}(\kappa)$ across WTP thresholds $\kappa$ (NMB in 100,000 USD/year). Center: posterior density of three-year restricted mean survival time (RMST; years). Right: posterior density of three-year mean total cost (MC; 100,000 USD). Dashed vertical lines denote posterior means.}
    \label{fig:posterior-results}
\end{figure}

Figure~\ref{fig:posterior-results} summarizes posterior inference.
The left panel reports posterior means and 95\% credible intervals for $\widehat{\Psi}_{d_1, d_2}(\kappa)$ over a range of $\kappa$.
At $\kappa = 0$ (a cost-only comparison), the NMB is near zero, indicating similar average costs under EBRT and VBT.
As $\kappa$ increases and survival receives greater weight, the posterior mean of NMB declines modestly; however, credible intervals include zero across most thresholds, providing limited evidence of a difference in cost-effectiveness.
The center and right panels show posterior densities for three-year restricted mean survival time~(RMST) and mean total cost~(MC), respectively.
RMST suggests a small survival advantage for VBT, while MC indicates slightly lower costs on average, with right-skewed tails reflecting infrequent high-cost events.
Since we did not restrict the cohort based on Medicare enrollment (continuous Part A and Part B without Part C), some claims may be missing.
This may result in the underestimation of total costs and should be considered when interpreting the findings.
Overall, the results do not indicate strong evidence of differential cost-effectiveness between EBRT and VBT over three years, though VBT may be favored due to modestly higher survival and lower cost.

\begin{figure}[tbp]
    \centering
    \includegraphics[width=0.5\linewidth]{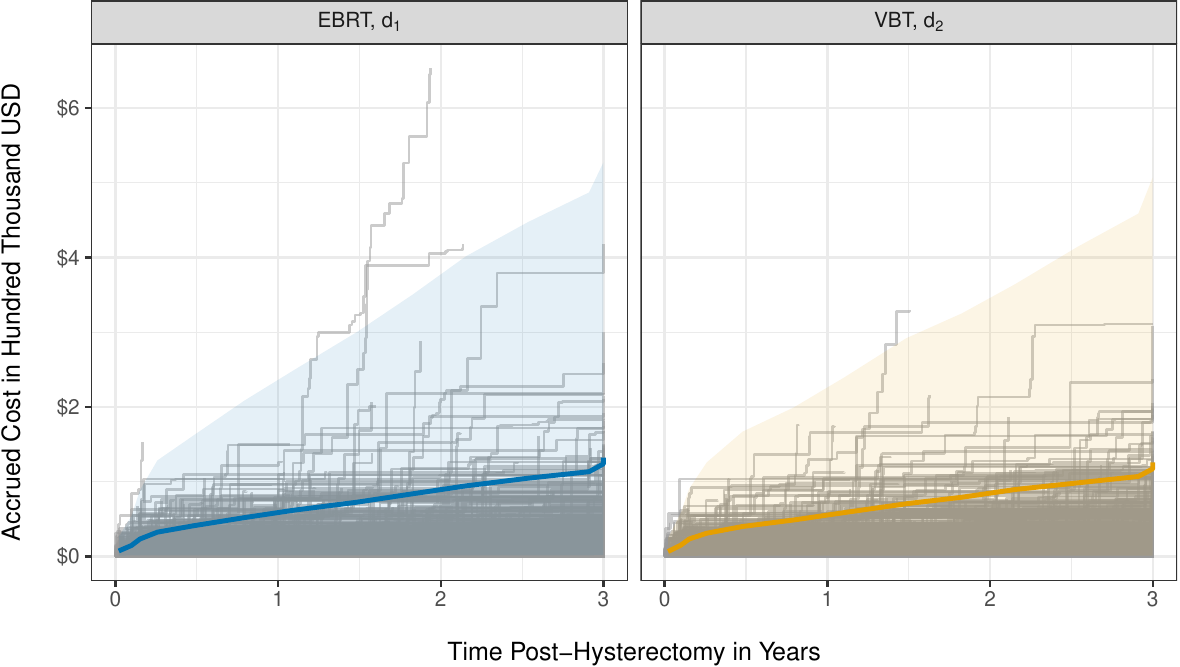}
    \caption{Posterior predictive cumulative cost trajectories for a representative stage IA patient under EBRT and VBT over three years. Thin gray curves are simulated trajectories from posterior draws; solid curves are posterior means with 95\% credible bands.}
    \label{fig:predictive-trajectories}
\end{figure}

To illustrate how the proposed approach supports posterior predictive visualization of counterfactual cost trajectories, Figure~\ref{fig:predictive-trajectories} shows examples for a representative stage IA patient (average age, white, urban residence, and low socioeconomic status).
Credible bands are computed using a binning approach: we partition the three-year horizon into 20 intervals and summarize simulated cumulative costs within each bin.
Under both strategies, most trajectories remain below \$100,000 over three years, and the posterior means are nearly indistinguishable, consistent with the NMB results in Figure~\ref{fig:posterior-results}.
EBRT shows slightly greater variability, with more trajectories exhibiting higher cost accrual, consistent with the broader dispersion in its MC distribution.

\section{Discussion}
\label{sec:closing}

We proposed a Bayesian joint modeling framework for CEA of time-varying treatments in observational claims data.
The approach jointly models gap times in continuous time and encounter-level costs, then uses Bayesian g-computation to obtain posterior inference for cost-effectiveness functionals such as MV and NMB.

In simulations, the proposed method performed comparably to correctly specified parametric models and was more robust than misspecified parametric and discrete-time comparators, particularly under substantial administrative censoring.
In the SEER-Medicare application to HR-EC, we found limited evidence of meaningful differences in three-year cost-effectiveness between adjuvant EBRT and VBT, with posterior summaries suggesting slightly lower costs and modestly higher survival under VBT.

This work has limitations.
First, identification relies on standard assumptions for longitudinal causal inference, including sequential ignorability and positivity; violations due to unmeasured confounding could bias causal contrasts.
Although we adjusted for a rich set of baseline and time-varying factors, these assumptions remain untestable.
Developing sensitivity analyses for time-varying confounding in continuous-time cost-effectiveness settings is an important direction for future work.
Second, although piecewise baseline functions reduce reliance on restrictive parametric forms, they introduce tuning choices, including the number and placement of time intervals.
% Finer partitions can reduce approximation bias but increase variance, especially when events are rare, late in follow-up.
Finer partitions can reduce approximation bias but increase variance, especially at later follow-up times when events are rare.
In our implementation, autoregressive priors on the piecewise constants (see Supplementary Section~\ref{supp:smoothing-priors}) encourage smoothness across adjacent intervals and improve stability, but additional work on adaptive partitioning~\citep{Li2021BayesianSurvival} or alternative priors could further reduce sensitivity to these choices.

Overall, the proposed framework provides a practical route to causal cost-effectiveness estimation under time-varying treatment, censoring, and irregular encounter times in claims data.
It supports a broad class of estimands expressed as functions of cumulative cost and survival, and yields uncertainty quantification through the posterior.
Future work could consider richer cost and survival models and develop sensitivity analyses for unmeasured time-varying confounding.

\subsection*{Acknowledgments}

We thank Colleen Brensinger and the staff of the Biostatistics Analysis Center, Center for Clinical Epidemiology and Biostatistics, Perelman School of Medicine, University of Pennsylvania, for assistance with preparation of the SEER-Medicare dataset.
This study used the linked SEER-Medicare database.
The interpretation and reporting of these data are the sole responsibility of the authors.
The authors acknowledge the efforts of the National Cancer Institute; Information Management Services (IMS), Inc.; and the Surveillance, Epidemiology, and End Results (SEER) Program tumor registries in the creation of the SEER-Medicare database.

The collection of cancer incidence data used in this study was supported by the California Department of Public Health pursuant to California Health and Safety Code Section 103885; Centers for Disease Control and Prevention's (CDC) National Program of Cancer Registries, under cooperative agreement NU58DP007156; the National Cancer Institute's Surveillance, Epidemiology, and End Results Program under contract HHSN261201800032I awarded to the University of California, San Francisco, contract HHSN261201800032I awarded to the University of Southern California, and contract HHSN261201800032I awarded to the Public Health Institute.
The ideas and opinions expressed herein are those of the authors and do not necessarily reflect the opinions of the State of California, Department of Public Health, the National Cancer Institute, and the Centers for Disease Control and Prevention or their Contractors and Subcontractors.

\subsection*{Conflict of Interest Disclosure}

The authors declare no conflicts of interest.

\subsection*{Data Availability Statement}

The analysis uses SEER-Medicare data provided by the National Cancer Institute Division of Cancer Control and Population Sciences, Healthcare Delivery Research Program.
Access to SEER-Medicare is governed by a Data Use Agreement; the data are not publicly available.

\subsection*{Funding Information}

This work was supported by the National Science Foundation Graduate Research Fellowship (Grant~2040433). 
Any opinions, findings, conclusions, or recommendations expressed in this material are those of the authors and do not necessarily reflect the views of the National Science Foundation.

\subsection*{Software and Code Availability}

Replication code and a sample dataset are available at \href{https://github.com/estfernan/causal-cea}{\texttt{github.com/estfernan/causal-cea}}.

\bibliography{references.bib}

@misc{Oganisian2020BayesianNonparametric,
	title = {Bayesian {Nonparametric} {Cost}-{Effectiveness} {Analyses}: {Causal} {Estimation} and {Adaptive} {Subgroup} {Discovery}},
	copyright = {arXiv.org perpetual, non-exclusive license},
	shorttitle = {Bayesian {Nonparametric} {Cost}-{Effectiveness} {Analyses}},
	url = {https://arxiv.org/abs/2002.04706},
	doi = {10.48550/ARXIV.2002.04706},
	urldate = {2025-12-09},
	publisher = {arXiv},
	author = {Oganisian, Arman and Mitra, Nandita and Roy, Jason},
	year = {2020},
	note = {arXiv:2002.04706},
}

@misc{Oganisian2024BayesianCounterfactual,
	title = {Bayesian {Counterfactual} {Prediction} {Models} for {HIV} {Care} {Retention} with {Incomplete} {Outcome} and {Covariate} {Information}},
	copyright = {Creative Commons Attribution 4.0 International},
	url = {https://arxiv.org/abs/2410.22481},
	doi = {10.48550/ARXIV.2410.22481},
	urldate = {2026-01-10},
	publisher = {arXiv},
	author = {Oganisian, Arman and Hogan, Joseph and Sang, Edwin and DeLong, Allison and Mosong, Ben and Fraser, Hamish and Mwangi, Ann},
	year = {2024},
	note = {arXiv:2410.22481},
}

@article{Suresh2022SurvivalPrediction,
	title = {Survival prediction models: an introduction to discrete-time modeling},
	volume = {22},
	issn = {1471-2288},
	shorttitle = {Survival prediction models},
	url = {https://bmcmedresmethodol.biomedcentral.com/articles/10.1186/s12874-022-01679-6},
	doi = {10.1186/s12874-022-01679-6},
	language = {en},
	number = {1},
	urldate = {2026-01-10},
	journal = {BMC Medical Research Methodology},
	author = {Suresh, Krithika and Severn, Cameron and Ghosh, Debashis},
	month = dec,
	year = {2022},
	pages = {207},
}

@article{Rubin1980RandomizationAnalysis,
	title = {Randomization {Analysis} of {Experimental} {Data}: {The} {Fisher} {Randomization} {Test} {Comment}},
	volume = {75},
	issn = {01621459},
	shorttitle = {Randomization {Analysis} of {Experimental} {Data}},
	url = {https://www.jstor.org/stable/2287653?origin=crossref},
	doi = {10.2307/2287653},
	number = {371},
	urldate = {2026-01-09},
	journal = {Journal of the American Statistical Association},
	author = {Rubin, Donald B.},
	month = sep,
	year = {1980},
	pages = {591},
}

@article{Li2021BayesianSurvival,
	title = {Bayesian survival analysis using gamma processes with adaptive time partition},
	volume = {91},
	issn = {0094-9655, 1563-5163},
	url = {https://www.tandfonline.com/doi/full/10.1080/00949655.2021.1912752},
	doi = {10.1080/00949655.2021.1912752},
	language = {en},
	number = {14},
	urldate = {2026-01-08},
	journal = {Journal of Statistical Computation and Simulation},
	author = {Li, Yi and Seo, Sumi and Lee, Kyu Ha},
	month = sep,
	year = {2021},
	pages = {2937--2952},
}

@article{Carpenter2017StanProbabilistic,
	title = {\textit{{Stan}} : {A} {Probabilistic} {Programming} {Language}},
	volume = {76},
	issn = {1548-7660},
	shorttitle = {\textit{{Stan}}},
	url = {http://www.jstatsoft.org/v76/i01/},
	doi = {10.18637/jss.v076.i01},
	language = {en},
	number = {1},
	urldate = {2026-01-04},
	journal = {Journal of Statistical Software},
	author = {Carpenter, Bob and Gelman, Andrew and Hoffman, Matthew D. and Lee, Daniel and Goodrich, Ben and Betancourt, Michael and Brubaker, Marcus and Guo, Jiqiang and Li, Peter and Riddell, Allen},
	year = {2017},
}

@article{Homan2014NoUturnSampler,
	title = {The {No}-{U}-turn sampler: adaptively setting path lengths in {Hamiltonian} {Monte} {Carlo}},
	volume = {15},
	issn = {1532-4435},
	number = {1},
	journal = {Journal of Machine Learning Research},
	author = {Homan, Matthew D. and Gelman, Andrew},
	month = jan,
	year = {2014},
	note = {Number of pages: 31
Publisher: JMLR.org
tex.issue\_date: January 2014},
	pages = {1593--1623},
}

@phdthesis{Watkins1989LearningDelayed,
	address = {Cambridge, UK},
	title = {Learning from {Delayed} {Rewards}},
	url = {http://www.cs.rhul.ac.uk/~chrisw/new_thesis.pdf},
	school = {King's College, University of Cambridge},
	author = {Watkins, Christopher John Cornish Hellaby},
	month = may,
	year = {1989},
}

@article{Robins1986NewApproach,
	title = {A new approach to causal inference in mortality studies with a sustained exposure period—application to control of the healthy worker survivor effect},
	volume = {7},
	copyright = {https://www.elsevier.com/tdm/userlicense/1.0/},
	issn = {02700255},
	url = {https://linkinghub.elsevier.com/retrieve/pii/0270025586900886},
	doi = {10.1016/0270-0255(86)90088-6},
	language = {en},
	number = {9-12},
	urldate = {2025-12-15},
	journal = {Mathematical Modelling},
	author = {Robins, James},
	year = {1986},
	pages = {1393--1512},
}

@article{Oganisian2024BayesianSemiparametric,
	title = {Bayesian semiparametric model for sequential treatment decisions with informative timing},
	volume = {25},
	copyright = {https://academic.oup.com/pages/standard-publication-reuse-rights},
	issn = {1465-4644, 1468-4357},
	url = {https://academic.oup.com/biostatistics/article/25/4/947/7560445},
	doi = {10.1093/biostatistics/kxad035},
	language = {en},
	number = {4},
	urldate = {2025-12-10},
	journal = {Biostatistics},
	author = {Oganisian, Arman and Getz, Kelly D and Alonzo, Todd A and Aplenc, Richard and Roy, Jason A},
	month = oct,
	year = {2024},
	pages = {947--961},
}

@techreport{NationalComprehensiveCancerNetwork2023NCCNClinical,
	address = {NCCN},
	title = {{NCCN} {Clinical} {Practice} {Guidelines} in {Oncology}: {Uterine} {Neoplasms}},
	url = {https://www.nccn.org},
	number = {Version 1.2024},
	institution = {NCCN},
	author = {{National Comprehensive Cancer Network}},
	month = sep,
	year = {2023},
}

@book{Varian1992MicroeconomicAnalysis,
	address = {New York, NY},
	edition = {3. ed., internat. student ed},
	title = {Microeconomic analysis},
	isbn = {978-0-393-95735-8 978-0-393-96026-6},
	language = {eng},
	publisher = {Norton},
	author = {Varian, Hal R.},
	year = {1992},
}

@book{Neumann2016CosteffectivenessHealth,
	address = {New York, NY},
	edition = {Second edition},
	title = {Cost-effectiveness in health and medicine},
	isbn = {978-0-19-049296-0},
	language = {eng},
	publisher = {Oxford University Press},
	editor = {Neumann, Peter and Ganiats, Theodore G. and Russell, Louise B. and Sanders, Gillian D. and Siegel, Joanna E.},
	year = {2016},
}

@article{Li2016PropensityScore,
	title = {Propensity score and doubly robust methods for estimating the effect of treatment on censored cost},
	volume = {35},
	copyright = {http://onlinelibrary.wiley.com/termsAndConditions\#vor},
	issn = {0277-6715, 1097-0258},
	url = {https://onlinelibrary.wiley.com/doi/10.1002/sim.6842},
	doi = {10.1002/sim.6842},
	language = {en},
	number = {12},
	urldate = {2025-12-09},
	journal = {Statistics in Medicine},
	author = {Li, Jiaqi and Handorf, Elizabeth and Bekelman, Justin and Mitra, Nandita},
	month = may,
	year = {2016},
	pages = {1985--1999},
}

@article{Nguyen2022GeneralFramework,
	title = {A general framework for optimising cost-effectiveness of pandemic response under partial intervention measures},
	volume = {12},
	issn = {2045-2322},
	url = {https://www.nature.com/articles/s41598-022-23668-x},
	doi = {10.1038/s41598-022-23668-x},
	language = {en},
	number = {1},
	urldate = {2025-12-09},
	journal = {Scientific Reports},
	author = {Nguyen, Quang Dang and Prokopenko, Mikhail},
	month = nov,
	year = {2022},
	pages = {19482},
}

@article{Li2018DoublyRobust,
	title = {A doubly robust approach for cost–effectiveness estimation from observational data},
	volume = {27},
	issn = {0962-2802, 1477-0334},
	url = {https://journals.sagepub.com/doi/10.1177/0962280217693262},
	doi = {10.1177/0962280217693262},
	language = {en},
	number = {10},
	urldate = {2025-12-09},
	journal = {Statistical Methods in Medical Research},
	author = {Li, Jiaqi and Vachani, Anil and Epstein, Andrew and Mitra, Nandita},
	month = oct,
	year = {2018},
	pages = {3126--3138},
}

@mastersthesis{Neyman1923ApplicationsTheorie,
	address = {Warsaw, Poland},
	title = {Sur les applications de la théorie des probabilités aux expériences agricoles: {Essai} des principes},
	language = {French},
	school = {University of Warsaw},
	author = {Neyman, Jerzy},
	year = {1923},
}

@article{Rubin1974EstimatingCausal,
	title = {Estimating causal effects of treatments in randomized and nonrandomized studies.},
	volume = {66},
	issn = {1939-2176, 0022-0663},
	url = {https://doi.apa.org/doi/10.1037/h0037350},
	doi = {10.1037/h0037350},
	language = {en},
	number = {5},
	urldate = {2025-12-16},
	journal = {Journal of Educational Psychology},
	author = {Rubin, Donald B.},
	month = oct,
	year = {1974},
	pages = {688--701},
}

@article{Wen2021ParametricGformula,
	title = {Parametric g‐formula implementations for causal survival analyses},
	volume = {77},
	issn = {0006-341X, 1541-0420},
	url = {https://academic.oup.com/biometrics/article/77/2/740-753/7445112},
	doi = {10.1111/biom.13321},
	language = {en},
	number = {2},
	urldate = {2025-12-15},
	journal = {Biometrics},
	author = {Wen, Lan and Young, Jessica G. and Robins, James M. and Hernán, Miguel A.},
	month = jun,
	year = {2021},
	pages = {740--753},
}

@article{Murphy2003OptimalDynamic,
	title = {Optimal {Dynamic} {Treatment} {Regimes}},
	volume = {65},
	copyright = {https://academic.oup.com/journals/pages/open\_access/funder\_policies/chorus/standard\_publication\_model},
	issn = {1369-7412, 1467-9868},
	url = {https://academic.oup.com/jrsssb/article/65/2/331/7092852},
	doi = {10.1111/1467-9868.00389},
	language = {en},
	number = {2},
	urldate = {2025-12-15},
	journal = {Journal of the Royal Statistical Society Series B: Statistical Methodology},
	author = {Murphy, S. A.},
	month = may,
	year = {2003},
	pages = {331--355},
}

@article{Xu2016BayesianNonparametric,
	title = {Bayesian {Nonparametric} {Estimation} for {Dynamic} {Treatment} {Regimes} {With} {Sequential} {Transition} {Times}},
	volume = {111},
	issn = {0162-1459, 1537-274X},
	url = {https://www.tandfonline.com/doi/full/10.1080/01621459.2015.1086353},
	doi = {10.1080/01621459.2015.1086353},
	language = {en},
	number = {515},
	urldate = {2025-12-15},
	journal = {Journal of the American Statistical Association},
	author = {Xu, Yanxun and Müller, Peter and Wahed, Abdus S. and Thall, Peter F.},
	month = jul,
	year = {2016},
	pages = {921--950},
}

@article{Spieker2020NestedComputation,
	title = {Nested \textit{g} -{Computation}: {A} {Causal} {Approach} to {Analysis} of {Censored} {Medical} {Costs} in {The} {Presence} of {Time}-{Varying} {Treatment}},
	volume = {69},
	copyright = {https://academic.oup.com/journals/pages/open\_access/funder\_policies/chorus/standard\_publication\_model},
	issn = {0035-9254, 1467-9876},
	shorttitle = {Nested \textit{g} -{Computation}},
	url = {https://academic.oup.com/jrsssc/article/69/5/1189/7058669},
	doi = {10.1111/rssc.12441},
	language = {en},
	number = {5},
	urldate = {2025-12-15},
	journal = {Journal of the Royal Statistical Society Series C: Applied Statistics},
	author = {Spieker, Andrew J. and Ko, Emily M. and Roy, Jason A. and Mitra, Nandita},
	month = nov,
	year = {2020},
	pages = {1189--1208},
}

@article{Spieker2018AnalyzingMedical,
	title = {Analyzing medical costs with time‐dependent treatment: {The} nested g‐formula},
	volume = {27},
	issn = {1057-9230, 1099-1050},
	shorttitle = {Analyzing medical costs with time‐dependent treatment},
	url = {https://onlinelibrary.wiley.com/doi/10.1002/hec.3651},
	doi = {10.1002/hec.3651},
	language = {en},
	number = {7},
	urldate = {2025-12-15},
	journal = {Health Economics},
	author = {Spieker, Andrew and Roy, Jason and Mitra, Nandita},
	month = jul,
	year = {2018},
	pages = {1063--1073},
}

@article{Clement2009UsingEffectiveness,
	title = {Using {Effectiveness} and {Cost}-effectiveness to {Make} {Drug} {Coverage} {Decisions}: {A} {Comparison} of {Britain}, {Australia}, and {Canada}},
	volume = {302},
	issn = {0098-7484},
	shorttitle = {Using {Effectiveness} and {Cost}-effectiveness to {Make} {Drug} {Coverage} {Decisions}},
	url = {http://jama.jamanetwork.com/article.aspx?doi=10.1001/jama.2009.1409},
	doi = {10.1001/jama.2009.1409},
	language = {en},
	number = {13},
	urldate = {2025-12-09},
	journal = {JAMA},
	author = {Clement, Fiona M. and Harris, Anthony and Li, Jing Jing and Yong, Karen and Lee, Karen M. and Manns, Braden J.},
	month = oct,
	year = {2009},
	pages = {1437},
}

@article{Avancena2021ExaminingEquity,
	title = {Examining {Equity} {Effects} of {Health} {Interventions} in {Cost}-{Effectiveness} {Analysis}: {A} {Systematic} {Review}},
	volume = {24},
	issn = {10983015},
	shorttitle = {Examining {Equity} {Effects} of {Health} {Interventions} in {Cost}-{Effectiveness} {Analysis}},
	url = {https://linkinghub.elsevier.com/retrieve/pii/S1098301520344545},
	doi = {10.1016/j.jval.2020.10.010},
	language = {en},
	number = {1},
	urldate = {2025-12-09},
	journal = {Value in Health},
	author = {Avanceña, Anton L.V. and Prosser, Lisa A.},
	month = jan,
	year = {2021},
	pages = {136--143},
}

@article{Handorf2019EstimatingCosteffectiveness,
	title = {Estimating cost-effectiveness from claims and registry data with measured and unmeasured confounders},
	volume = {28},
	issn = {0962-2802, 1477-0334},
	url = {https://journals.sagepub.com/doi/10.1177/0962280218759137},
	doi = {10.1177/0962280218759137},
	language = {en},
	number = {7},
	urldate = {2025-12-09},
	journal = {Statistical Methods in Medical Research},
	author = {Handorf, Elizabeth A and Heitjan, Daniel F and Bekelman, Justin E and Mitra, Nandita},
	month = jul,
	year = {2019},
	pages = {2227--2242},
}

@article{Hua2022PersonalizedDynamic,
	title = {Personalized {Dynamic} {Treatment} {Regimes} in {Continuous} {Time}: {A} {Bayesian} {Approach} for {Optimizing} {Clinical} {Decisions} with {Timing}},
	volume = {17},
	issn = {1936-0975},
	shorttitle = {Personalized {Dynamic} {Treatment} {Regimes} in {Continuous} {Time}},
	url = {https://projecteuclid.org/journals/bayesian-analysis/volume-17/issue-3/Personalized-Dynamic-Treatment-Regimes-in-Continuous-Time--A-Bayesian/10.1214/21-BA1276.full},
	doi = {10.1214/21-BA1276},
	number = {3},
	urldate = {2025-12-15},
	journal = {Bayesian Analysis},
	author = {Hua, William and Mei, Hongyuan and Zohar, Sarah and Giral, Magali and Xu, Yanxun},
	month = sep,
	year = {2022},
}

@article{Chakraborty2014DynamicTreatment,
	title = {Dynamic {Treatment} {Regimes}},
	volume = {1},
	issn = {2326-8298, 2326-831X},
	url = {https://www.annualreviews.org/doi/10.1146/annurev-statistics-022513-115553},
	doi = {10.1146/annurev-statistics-022513-115553},
	language = {en},
	number = {1},
	urldate = {2025-12-15},
	journal = {Annual Review of Statistics and Its Application},
	author = {Chakraborty, Bibhas and Murphy, Susan A.},
	month = jan,
	year = {2014},
	pages = {447--464},
}

@article{Illenberger2023IdentifyingOptimally,
	title = {Identifying optimally cost-effective dynamic treatment regimes with a {Q}-learning approach},
	volume = {72},
	copyright = {https://academic.oup.com/pages/standard-publication-reuse-rights},
	issn = {0035-9254, 1467-9876},
	url = {https://academic.oup.com/jrsssc/article/72/2/434/7071570},
	doi = {10.1093/jrsssc/qlad016},
	language = {en},
	number = {2},
	urldate = {2025-12-15},
	journal = {Journal of the Royal Statistical Society Series C: Applied Statistics},
	author = {Illenberger, Nicholas and Spieker, Andrew J and Mitra, Nandita},
	month = may,
	year = {2023},
	pages = {434--449},
}

@article{Chen2025FlexibleBayesian,
	title = {A flexible {Bayesian} g-formula for causal survival analyses with time-dependent confounding},
	volume = {31},
	issn = {1380-7870, 1572-9249},
	url = {https://link.springer.com/10.1007/s10985-025-09652-3},
	doi = {10.1007/s10985-025-09652-3},
	language = {en},
	number = {2},
	urldate = {2025-12-15},
	journal = {Lifetime Data Analysis},
	author = {Chen, Xinyuan and Hu, Liangyuan and Li, Fan},
	month = apr,
	year = {2025},
	pages = {394--421},
}

@article{Baio2014BayesianModels,
	title = {Bayesian models for cost‐effectiveness analysis in the presence of structural zero costs},
	volume = {33},
	copyright = {http://creativecommons.org/licenses/by-nc-nd/3.0/},
	issn = {0277-6715, 1097-0258},
	url = {https://onlinelibrary.wiley.com/doi/10.1002/sim.6074},
	doi = {10.1002/sim.6074},
	language = {en},
	number = {11},
	urldate = {2025-12-15},
	journal = {Statistics in Medicine},
	author = {Baio, Gianluca},
	month = may,
	year = {2014},
	pages = {1900--1913},
}

\subsection*{Supporting Information}

Supporting information is available online with this article.

\clearpage
\newpage

\appendix

\section{Identification Assumptions}
\label{supp:identification-assumptions}

We state the assumptions required to identify $\psi_{\bsd}(\kappa)$ and $\psi_{\bsd}(g)$ 
from the observed data.
Throughout, we use $\bsH_j$ to denote the observed history available at encounter $j$
(defined in the main text),
and $\cF(\bsH_j) \subseteq \cA$ to denote the feasible action set at history $\bsH_j$.
We adopt the event coding $\delta_j \in \qty{0, 1, 2}$ 
for censoring, transition to a new encounter, and death, respectively.

\begin{assumption}[Stable unit treatment value assumption]
    Each patient's potential outcomes depend solely on their treatment history (no interference),
    and the observed outcomes are equal to the corresponding potential outcomes under 
    the observed treatment history (consistency).
    Formally, for any encounter $j$: 
    if $\obar{A}_{ij} = \obar{a}_j$, then
    \begin{align*}
        & \qty(Y_{ij}, W_{i(j+1)}, \delta_{i(j+1)}, \bsL_{i(j+1)}, Z_{i(j+1)}) \\
        & \qquad = \qty(Y_{ij}(\obar{a}_j), W_{i(j+1)}(\obar{a}_j), \delta_{i(j+1)}(\obar{a}_j), \bsL_{i(j+1)}(\obar{a}_j), Z_{i(j+1)}(\obar{a}_j)),
    \end{align*}
    whenever these quantities are well-defined (that is, before terminal events).
\end{assumption}

\begin{assumption}[Treatment positivity]
    Every treatment option that is feasible at a given history occurs with 
    positive probability in the observed data. 
    Formally, for each encounter $j$ and every history vector $\bsh_j$ with $f(\bsh_j) > 0$, 
    we have
    \begin{equation*}
        \Pr(A_j = a_j \mid \bsH_j = \bsh_j) > 0, \quad
        \text{for all} \ a_j \in \cF(\bsh_j).
    \end{equation*}
\end{assumption}

\begin{assumption}[Sequential ignorability]
    Conditional on the observed history, treatment decisions are as if randomized:
    there are no unmeasured variables that jointly affect treatment and 
    the future counterfactual trajectory.
    Formally, for each encounter $j$ and each treatment history $\obar{a}_j$,
    \begin{equation*}
        \qty(
            \ubar{Y}_j(\obar{a}_j), \ubar{W}_{j+1}(\obar{a}_j), \ubar{\delta}_{j+1}(\obar{a}_j), 
            \ubar{\bsL}_{j+1}(\obar{a}_j), \ubar{Z}_{j+1}(\obar{a}_j)
        ) \indep A_j \mid \bsH_j.
    \end{equation*}
\end{assumption}

\begin{assumption}[Non-informative censoring given observed history]
    Censoring may depend on the observed history, but it does not 
    depend on counterfactual outcomes.
    We state this in terms of the censoring hazard.
    Formally, let $h_C(u \mid \bsS_j = \bss_j)$ denote the cause-specific hazard of censoring at elapsed time $u$ following the $(j-1)$-th encounter, conditional on the history before encounter $j$.
    Then, for each $j$, each $u \geq 0$, and every treatment history $\obar{a}_j$,
    \begin{equation*}
        h_C\qty(
            u \mid \bsS_j = \bss_j, 
            \ubar{Y}_j(\obar{a}_j), \ubar{W}_{j+1}(\obar{a}_j), \ubar{\delta}_{j+1}(\obar{a}_j),
            \ubar{\bsL}_{j+1}(\obar{a}_j), \ubar{Z}_{j+1}(\obar{a}_j)
        ) = h_C\qty(u \mid \bsS_j = \bss_j),
    \end{equation*}
    for every $\bsh_j$.
\end{assumption}

\begin{assumption}[Censoring positivity]
    For trajectories that occur under the treatment history (or regimes) of interest, 
    there is a positive probability of remaining uncensored long enough to contribute information 
    (up to the relevant horizon, if applicable).
    Formally, let $S_C(u \mid \bsS_j = \bss_j)$ be the conditional censoring survival function
    \begin{equation*}
        S_C(u \mid \bsS_j = \bss_j)
            = \exp{-\int_0^u h_C(v \mid \bsS_j = \bss_j) \, \dd v}.
    \end{equation*}
    For a fixed horizon $\tau$ (if used), assume that for each encounter $j$ and every history $\bsh_j$ with positive density under the treatment histories (or regimes) being compared,
    \begin{equation*}
        S_C(u \mid \bsS_j = \bss_j) > 0 \quad
            \text{for all } u \in [0, \tau].
    \end{equation*}
\end{assumption}

\section{Proof of g-Formula}
\label{supp:g-formula}

Estimation of $\psi_{\bsd}(g)$ requires the joint distribution of $T(\bsd)$ and $U(\bsd)$.
It suffices to characterize the joint distribution of the counterfactual trajectory 
$\qty(\obar{Y}(\bsd), \obar{W}(\bsd), J(\bsd))$, since
$T(\bsd) = \sum_{s=1}^{J(\bsd)} W_s(\bsd)$ and $U(\bsd) = \sum_{s=1}^{J(\bsd)} Y_s(\bsd)$.
Moreover, fixing $J(\bsd) = r$ determines the terminal transition pattern 
$\obar{\delta}(\bsd) = (\obar{1}_{r-1}, 2)$, which we denote by
$\bse_r \coloneqq (\obar{1}_{r-1}, 2)$.
Therefore, summing over the possible values of $J(\bsd)$, identification of $\psi_{\bsd}(g)$
rests on identifying the joint distribution of
$\qty(\obar{Y}(\bsd), \obar{W}(\bsd), \obar{\delta}(\bsd))$ under the standard assumptions 
in Supplementary Section~\ref{supp:identification-assumptions}.

To make the derivation explicit, we first establish the result for an arbitrary treatment history
$\obar{a}_r = (a_1, \ldots, a_r)$.
The corresponding counterfactual density is
$f\qty(\obar{Y}(\obar{a}_r), \obar{W}(\obar{a}_r), \obar{\delta}(\obar{a}_r))$.
A dynamic treatment regime $\bsd$ induces a history through $a_j = d_j(\bsH_j)$, so the same
derivation applies under $\bsd$ by substituting $\obar{a}_r = \obar{a}^{\bsd}_r$.

Here we provide the derivation for $r = 2$, noting that the general case $r > 2$ follows by
iterative integration.
For simplicity, we omit baseline covariates $\bsL_0$, which can be incorporated by
integrating with respect to $f(\bsL_0)$ at the beginning of the proof.

For $r = 2$, we have $\bse_2 = (1, 2)$, and the joint density can be written as
\begin{align*}
    & f\qty(\obar{Y}(\obar{a}_2) = \obar{y}, \obar{W}(\obar{a}_2) = \obar{w}, \obar{\delta}(\obar{a}_2) = \bse_2) \\
    & \quad = f\qty(Y_1(\obar{a}_1) = y_1, Y_2(\obar{a}_2) = y_2, W_1(\obar{a}_0) = w_1, W_2(\obar{a}_1) = w_2, \delta_1(\obar{a}_0) = 1, \delta_2(\obar{a}_1) = 2) \\
    & \quad = f\qty(Y_1(\obar{a}_1) = y_1, Y_2(\obar{a}_2) = y_2, W_1 = w_1, W_2(\obar{a}_1) = w_2, \delta_1 = 1, \delta_2(\obar{a}_1) = 2) \\
    & \quad = f\qty(Y_1(\obar{a}_1) = y_1, Y_2(\obar{a}_2) = y_2, W_2(\obar{a}_1) = w_2, \delta_2(\obar{a}_1) = 2 \mid W_1 = w_1, \delta_1 = 1) f(W_1 = w_1, \delta_1 = 1),
\end{align*}
where $\obar{a}_0 = 0$ by design.
Note that $W_1$ and $\delta_1$ are observed before any post-surgery treatment decision,
so they do not depend on a treatment assignment and can be factored into a separate density.

To connect counterfactual outcomes with the observed data $\bsO$, we first define
$\bsX_j = (\bsL_j, Z_j) \in \cX_j$ as the covariates and treatment-readiness indicator, with 
$\cX_j \coloneq \cL \times \qty{0, 1}$.
Integrating over the first set of confounders yields
\begin{align*}
    & f\qty(Y_1(\obar{a}_1) = y_1, Y_2(\obar{a}_2) = y_2, W_2(\obar{a}_1) = w_2, \delta_2(\obar{a}_1) = 2 \mid W_1 = w_1, \delta_1 = 1) \\
    & \quad = \int_{\cX_1} f\qty(Y_1(\obar{a}_1) = y_1, Y_2(\obar{a}_2) = y_2, W_2(\obar{a}_1) = w_2, \delta_2(\obar{a}_1) = 2 \mid \bsX_1 = \bsx_1, W_1 = w_1, \delta_1 = 1) \\
    & \qquad \qquad \times f\qty(\bsX_1 = \bsx_1 \mid W_1 = w_1, \delta_1 = 1) \, \dd \bsx_1 \\
    & \quad = \int_{\cX_1} f\qty(Y_1(\obar{a}_1) = y_1, Y_2(\obar{a}_2) = y_2, W_2(\obar{a}_1) = w_2, \delta_2(\obar{a}_1) = 2 \mid \bsH_1 = \bsh_1) \\
    & \qquad \qquad \times f\qty(\bsX_1 = \bsx_1 \mid W_1 = w_1, \delta_1 = 1) \, \dd \bsx_1,
\end{align*}
where $\bsH_1 = (\bsX_1, W_1, \delta_1)$ is the observed history at $j = 1$.

Under treatment positivity and sequential ignorability,
\(
    Y_1(\obar{a}_1), Y_2(\obar{a}_2), W_2(\obar{a}_1), \delta_2(\obar{a}_1) \indep A_1 \mid \bsH_1.
\)
Therefore, conditioning on $\bsH_1$, we may introduce $A_1$ and then apply consistency; 
that is, the consistency component of SUTVA~\citep{Rubin1980RandomizationAnalysis},
which gives
\begin{align*}
    & f\qty(Y_1(\obar{a}_1) = y_1, Y_2(\obar{a}_2) = y_2, W_2(\obar{a}_1) = w_2, \delta_2(\obar{a}_1) = 2 \mid \bsH_1 = \bsh_1) \\
    & \quad = f\qty(Y_1(\obar{a}_1) = y_1, Y_2(\obar{a}_2) = y_2, W_2(\obar{a}_1) = w_2, \delta_2(\obar{a}_1) = 2 \mid A_1 = a_1, \bsH_1 = \bsh_1) \\
    & \quad = f\qty(Y_1 = y_1, Y_2(\obar{a}_2) = y_2, W_2 = w_2, \delta_2 = 2 \mid A_1 = a_1, \bsH_1 = \bsh_1).
\end{align*}
Factoring the conditional density yields
\begin{align*}
    & f\qty(Y_1 = y_1, Y_2(\obar{a}_2) = y_2, W_2 = w_2, \delta_2 = 2 \mid A_1 = a_1, \bsH_1 = \bsh_1) \\
    & \quad = f\qty(Y_2(\obar{a}_2) = y_2 \mid W_2 = w_2, \delta_2 = 2, Y_1 = y_1, A_1 = a_1, \bsH_1 = \bsh_1) \\
    & \qquad \qquad \times f\qty(W_2 = w_2, \delta_2 = 2 \mid Y_1 = y_1, A_1 = a_1, \bsH_1 = \bsh_1)
        f\qty(Y_1 = y_1 \mid A_1 = a_1, \bsH_1 = \bsh_1) \\
    & \quad = f\qty(Y_2(\obar{a}_2) = y_2 \mid W_2 = w_2, \delta_2 = 2, \bsS_2 = \bss_2) \\
    & \qquad \qquad \times f\qty(W_2 = w_2, \delta_2 = 2 \mid \bsS_2 = \bss_2) 
        f\qty(Y_1 = y_1 \mid A_1 = a_1, \bsH_1 = \bsh_1),
\end{align*}
where $\bsS_2 = (Y_1, A_1, \bsH_1)$ denotes the data observed through encounter $j = 1$.

Next, we integrate over the second set of confounders $\bsX_2 = (\bsL_2, Z_2)$ with support $\cX_2$ to evaluate 
the density of $Y_2(\obar{a}_2)$:
\begin{align*}
    & f\qty(Y_2(\obar{a}_2) = y_2 \mid W_2 = w_2, \delta_2 = 2, \bsS_2 = \bss_2) \\
    & \quad = \int_{\cX_2} f\qty(Y_2(\obar{a}_2) = y_2 \mid \bsX_2 = \bsx_2, W_2 = w_2, \delta_2 = 2, \bsS_2 = \bss_2) \\
    & \qquad \qquad \times f\qty(\bsX_2 = \bsx_2 \mid W_2 = w_2, \delta_2 = 2, \bsS_2 = \bss_2) \, \dd \bsx_2 \\
    & \quad = \int_{\cX_2} f\qty(Y_2(\obar{a}_2) = y_2 \mid \bsH_2 = \bsh_2) \\
    & \qquad \qquad \times f\qty(\bsX_2 = \bsx_2 \mid W_2 = w_2, \delta_2 = 2, \bsS_2 = \bss_2) \, \dd \bsx_2,
\end{align*}
where $\bsH_2 = (\bsX_2, W_2, \delta_2, \bsS_2)$ is the history at $j = 2$.

By sequential ignorability, $Y_2(\obar{a}_2) \indep A_2 \mid \bsH_2$, 
and by consistency, $Y_2(\obar{a}_2) = Y_2$ when $\obar{A}_2 = \obar{a}_2$.
Thus,
\begin{align*}
    & f\qty(Y_2(\obar{a}_2) = y_2 \mid \bsH_2 = \bsh_2) \\
    & \quad = f\qty(Y_2(\obar{a}_2) = y_2 \mid A_2 = a_2, \bsH_2 = \bsh_2) \\
    & \quad = f\qty(Y_2 = y_2 \mid A_2 = a_2, \bsH_2 = \bsh_2).
\end{align*}
Combining all pieces, we obtain the joint density of the counterfactual outcomes:
\begin{align*}
    & f\qty(\obar{Y}(\bsd) = \obar{y}, \obar{W}(\bsd) = \obar{w}, \obar{\delta}(\bsd) = \bse_r) \\
    & \quad = \int_{\cX_1} \int_{\cX_2} f\qty(Y_2 = y_2 \mid A_2 = a_2, \bsH_2 = \bsh_2) \\
    & \qquad \qquad \times f\qty(\bsX_2 = \bsx_2 \mid W_2 = w_2, \delta_2 = 2, \bsS_2 = \bss_2) 
        f\qty(W_2 = w_2, \delta_2 = 2 \mid \bsS_2 = \bss_2) \\
    & \qquad \qquad \times f\qty(Y_1 = y_1 \mid A_1 = a_1, \bsH_1 = \bsh_1) \\
    & \qquad \qquad \times f\qty(\bsX_1 = \bsx_1 \mid W_1 = w_1, \delta_1 = 1)
        f\qty(W_1 = w_1, \delta_1 = 1) \, \dd \bsx_2 \, \dd \bsx_1.
\end{align*}

Extensions to $r > 2$ follow by repeating iterative integration and the application of 
sequential ignorability and consistency.
In the main text, we specify parametric or semiparametric models for each conditional density and 
develop a posterior sampling algorithm that uses g-computation~\citep{Robins1986NewApproach} 
to estimate counterfactual estimands $\psi_{\bsd}(\kappa)$, $\psi_{\bsd}(g)$, and contrasts such as
$\Psi_{\bsd, \bsd'}(\kappa)$.~$\qed$

\section{Likelihood Contributions}
\label{supp:likelihood-contributions}

We construct the likelihood within a standard longitudinal framework 
under non-informative censoring.
Let $\bstheta$ denote the full parameter vector indexing the joint models, 
\begin{equation*}
    \bstheta 
        = (
            \bsh_{01}, \bsvarphi_1, \bsh_{02}, \bsvarphi_2,
            \bseta, \bsphi, 
            \bsm_0, \bsbeta, \zeta
        ).
\end{equation*}
For each patient $i$, the contribution of the observed data $\bsO_i$ to the likelihood, 
conditional on $\bstheta$, is
\begin{align*}
    & f(\bsO_i \mid \bstheta) 
        = S(w_{iJ_i} \mid \bsS_{J_i} = \bss_{iJ_i}, \bsh_{01}, \bsvarphi_1, \bsh_{02}, \bsvarphi_2) 
            h(w_{iJ_i}, \delta_{iJ_i} = 2 \mid \bsS_{J_i} = \bss_{iJ_i}, \bsh_{02}, \bsvarphi_2)^{\one{\delta_{iJ_i} = 2}} \\
    & \qquad \times \prod_{j=1}^{J_i} 
        f(y_{ij} \mid A_j = a_{ij}, \bsH_j = \bsh_{ij}, \bsm_0, \bsbeta, \zeta) \\
    & \qquad \times \prod_{j=1}^{J_i} 
        f(z_{ij} \mid \bsL_j = \bsl_{ij}, W_j = w_{ij}, \delta_j = \delta_{ij}, \bsS_j = \bss_{ij}, \bsphi) \\
    & \qquad \times \prod_{j=1}^{J_i} 
        f(\bsl_{ij} \mid W_j = w_{ij}, \delta_j = \delta_{ij}, \bsS_j = \bss_{ij}, \bseta) \\
    & \qquad \times \prod_{j=1}^{J_i-1} 
        S(w_{ij} \mid \bsS_j = \bss_{ij}, \bsh_{01}, \bsvarphi_1, \bsh_{02}, \bsvarphi_2)
        h(w_{ij}, \delta_{ij} = 1 \mid \bsS_j = \bss_{ij}, \bsh_{01}, \bsvarphi_1),
\end{align*}
where $S(\cdot)$ denotes the overall survival function defined in the main text.

Patients contribute to the likelihood across all observed encounters.
At the last observed index $j = J_i$, censored individuals ($\delta_{iJ_i} = 0$) 
contribute only through the survival term 
$S(w_{iJ_i} \mid \bsS_{iJ_i})$, whereas deaths ($\delta_{iJ_i} = 2$) 
contribute through both $S(w_{iJ_i} \mid \bsS_{iJ_i})$ and 
the death-specific hazard term $h(w_{iJ_i}, \delta_{iJ_i} = 2 \mid \bsS_{J_i} = \bss_{iJ_i})$.
Aggregating over all $n$ patients, the full likelihood is 
\begin{equation*}
    f(\bsO \mid \bstheta) = \prod_{i=1}^n f(\bsO_i \mid \bstheta).
\end{equation*}

\section{Modeling Covariate Process and Treatment Readiness}
\label{supp:covariates-treatment-readiness}

From the g-formula, the joint conditional distribution of the covariates and 
the treatment-readiness indicator $\bsX_j \coloneqq (\bsL_j, Z_j)$ can be decomposed as
\begin{align*}
    & f(\bsx_j \mid W_j = w_j, \delta_j = k, \bsS_j = \bss_j) \\
    & \quad = f(z_j \mid \bsL_j = \bsl_j, W_j = w_j, \delta_j = k, \bsS_j = \bss_j)
        f(\bsl_j \mid W_j = w_j, \delta_j = k, \bsS_j = \bss_j).
\end{align*}
For tractability, we assume the components of 
$\bsL_j = (L_{j1}, \ldots, L_{jP})$ 
are conditionally independent given the conditioning terms.
We then fit separate regression models appropriate to the support of each component.

Let $\cP_0 \subseteq \qty{1, \ldots, P}$ index the baseline components of $\bsL_j$, 
denoted $\bsL_0$, that are measured at surgery and 
treated as time-invariant patient characteristics.
For each $p \in \cP_0$, we model $L_{0p}$ using a simple parametric distribution 
(for example, normal for continuous variables and Bernoulli for binary variables), 
with likelihood contributions based only on the value observed at surgery in $\bsO$.

Let $\cP_1 \coloneqq \qty{1, \ldots, P} \setminus \cP_0$ 
index the remaining covariates that may evolve over encounters 
(that is, time-varying characteristics).
We model each $L_{jp}$ for $p \in \cP_1$ using a generalized linear model 
conditional on the most recent observed state 
$\bsQ_j \coloneqq (W_j, \delta_j, \bsS_j)$:
\begin{equation*}
    g_p\qty(\E{L_{jp}}{\bsQ_j = \bsq_j}) = \bsomega_{0p} + \bsq_j' \bsomega_p,
\end{equation*}
where $g_p$ is a link function chosen to match the support of $L_{jp}$ 
(for example, identity for continuous variables and logit for binary variables), 
$\omega_{0p}$ is an intercept, and $\bsomega_p$ is a coefficient vector.
Although conditional independence is assumed for simplicity, 
the covariate model could be generalized to a multivariate specification 
if modeling correlations is important.
Recall that, in our SEER-Medicare application, 
we specify time-homogeneity and first-order Markov restrictions for $\bsS_j$.

We treat $Z_j$ as a discrete time-to-event indicator with a monotone trajectory: 
once $Z_j = 1$, then $Z_k = 1$ for all $k > j$.
Accordingly, we use a discrete-time hazard (continuation-ratio) model 
for treatment readiness:
\begin{equation*}
    \Pr(Z_j = 1 \mid \bsL_j = \bsl_j, \bsQ_j = \bsq_j, Z_{j-1} = 0) 
        = \logit^{-1}\qty(\phi_j + \bsl_j' \bsphi_L + \bsq_j' \bsphi_Q),
\end{equation*}
where $\phi_j$ is the encounter-specific baseline logit-hazard, 
$\bsphi_L$ captures covariate effects, and 
$\bsphi_Q$ captures dependence on the recent state;
see~\citet{Suresh2022SurvivalPrediction} for further details.
Only the sequence up to and including the first $Z_j = 1$ contributes to the likelihood; 
for example, if readiness starts at $j = 3$, 
then $(Z_1 = 0, Z_2 = 0, Z_3 = 1)$ informs the model, and 
later values are excluded.

Let $\bseta \coloneqq (\bseta_{L_0}, \bseta_{L_1})$ 
denote the covariate-model parameters, where
$\bseta_{L_0} \coloneqq \qty{\bsgamma_p \colon p \in \cP_0}$ and 
$\bseta_{L_1} \coloneqq \qty{(\omega_{0p}, \bsomega_p) \colon p \in \cP_1}$.
Here, $\bsgamma_p$ denotes the parameter(s) indexing the baseline marginal model for $L_{0p}$.
Let $\bsphi \coloneqq (\phi_1, \ldots, \phi_N, \bsphi_L, \bsphi_Q)$ 
denote the readiness-model parameters, where 
$N \coloneqq \max[i \leq n] J_i$ is the maximum number of observed encounters.
For example, baseline age might be modeled as 
$L_{01} \sim \Normal(\mu_{01}, \sigma_{01}^2)$ 
with $\bsgamma_1 = (\mu_{01}, \sigma_{01})$, and baseline marital status as 
$L_{02} \sim \Bern(\vartheta_{03})$ 
with $\bsgamma_2 \equiv \vartheta_{02}$.

\section{Prior Specification and Posterior Factorization}
\label{supp:implementation-details}

Assuming prior independence across components of $\bstheta$, 
the joint prior factorizes as
\begin{equation*}
    f(\bstheta) 
        = f(\bsh_{01}) f(\bsvarphi_1) 
            f(\bsh_{02}) f(\bsvarphi_2) 
            f(\bseta) f(\bsphi) 
            f(\bsm_0) f(\bsbeta) f(\zeta),
\end{equation*}
where $\bseta$ and $\bsphi$ denote the covariate and treatment-readiness model parameters 
(Supplementary Section~\ref{supp:covariates-treatment-readiness}).
The posterior distribution is
\begin{equation*}
    f(\bstheta \mid \bsO) 
        \propto f(\bsO \mid \bstheta) f(\bstheta),
\end{equation*}
where $f(\bsO \mid \bstheta)$ is introduced in Supplementary Section~\ref{supp:likelihood-contributions}.
The component likelihoods (for example, Bernoulli, gamma, and normal) 
are substituted directly in our \texttt{Stan} 
implementation~\citep{Carpenter2017StanProbabilistic}.
Posterior sampling yields draws $\bstheta^{(1)}, \ldots, \bstheta^{(M)}$, 
which are used for inference in the main text.

We assign weakly informative priors so that, in moderate to large samples, 
posterior inference is driven primarily by the likelihood.
For baseline covariates measured at surgery, we use standard normal and Bernoulli priors; 
for example, 
\begin{gather*}
    \mu_{0p} \sim \Normal(0, 1), \quad \sigma_{0p} \sim \HN(1), \\
    \vartheta_{0p} \sim \Unif(0, 1).
\end{gather*}
All regression coefficients---including those in the gap time, covariate, 
readiness, and cost models---follow independent normal priors $\Normal(0, 3^2)$ 
(or the multivariate analog $\MVN(\mathbf{0}, 3^2 \bsI)$):
\begin{gather*}
    \bsvarphi_k \sim \MVN(\mathbf{0}_{P_S}, 3^2 \bsI_{P_S}), \quad k \in \qty{1, 2}, \\
    \beta_k \sim \Normal(0, 3^2), \quad k \in \qty{1, 2}, \\
    \bsbeta_A \sim \MVN(\mathbf{0}_{P_A}, 3^2 \bsI_{P_A}), \quad
    \bsbeta_H \sim \MVN(\mathbf{0}_{P_H}, 3^2 \bsI_{P_H}), \\
    \omega_{0p} \sim \Normal(0, 3^2), \quad
    \bsomega_p \sim \MVN(\mathbf{0}_{P_X}, 3^2 \bsI_{P_X}), \quad p \in \cP_1, \\
    \bsphi_L \sim \MVN(\mathbf{0}_{P_L}, 3^2 \bsI_{P_L}), \quad
    \bsphi_X \sim \MVN(\mathbf{0}_{P_X}, 3^2 \bsI_{P_X}),
\end{gather*}
where $P_S$, $P_A$, $P_H$, $P_L$, and $P_X$ denote the corresponding vector lengths.
Finally, the cost-model variance parameter follows a half-normal prior, $\zeta \sim \HN(1)$.

Priors for the piecewise baseline functions (that is, baseline hazards and mean costs) 
use first-order autoregressive smoothing across intervals and 
are given in Supplementary Section~\ref{supp:smoothing-priors}.

\section{Autoregressive Smoothing Priors for Baseline Functions}
\label{supp:smoothing-priors}

Baseline mean costs $\qty{\log m_{0q}}_{q=1}^Q$ and 
cause-specific hazard rates $\qty{\log h_{0kq}}_{q=1}^Q$ are assigned 
first-order autoregressive (AR(1)) priors on the log scale across $Q$ time intervals.
We denote this AR(1) prior by $\AR(\mu, \rho, \sigma)$; 
see~\citet{Oganisian2024BayesianCounterfactual,Oganisian2024BayesianSemiparametric} 
for further details.
Specifically,
\begin{gather*}
    \log m_{0q} \sim \AR(m_0, \rho_m, \sigma_m), \\
    \log h_{0kq} \sim \AR(h_{0k}, \rho_{h_k}, \sigma_{h_k}), \quad 
    k \in \qty{1, 2}.
\end{gather*}
The $\AR(\mu, \rho, \sigma)$ prior is defined recursively as
\begin{gather*}
    \log m_{01} = m_0 + \sigma_m \varepsilon_{m_1}, \\
    \log m_{0q} = m_0 (1 - \rho_m) + \rho_m \log m_{0(q-1)} + \sigma_m \varepsilon_{m_q}, \quad 
    q = 2, \ldots, Q,
\end{gather*}
where $\varepsilon_{m_q} \iid \Normal(0, 1)$.
This construction implies
\begin{equation*}
    \E{\log m_{0q}} = m_0, \quad 
    \Corr{\log m_{0q}}{\log m_{0(q-1)}} = \rho_m, \quad
    \Var{\log m_{0q}} = \frac{\sigma_m^2}{1 - \rho_m^2}.
\end{equation*}
Analogous expressions hold for $\qty{\log h_{0kq}}_{q=1}^Q$ 
with parameters $(h_{0k}, \rho_{h_k}, \sigma_{h_k})$.
We assign weakly informative priors to the AR(1) hyperparameters:
\begin{gather*}
    m_0 \sim \Normal(0, 1), \\
    h_{0k} \sim \Normal(0, 1), \quad 
    k \in \qty{1, 2}, \\
    \rho_m = 2 \logit^{-1}(\widetilde{\rho}_m) - 1, \quad 
    \widetilde{\rho}_m \sim \BetaDist(2, 2) \\
    \rho_{h_k} = 2 \logit^{-1}(\widetilde{\rho}_{h_k}) - 1, \quad
    \widetilde{\rho}_{h_k} \sim \BetaDist(2, 2), \quad 
    k \in \qty{1, 2}, \\
    \sigma_m \sim \HN(1), \\
    \sigma_{h_k} \sim \HN(1), \quad 
    k \in \qty{1, 2}.
\end{gather*}

Finally, the encounter-specific intercepts $\qty{\phi_j}_{j=1}^N$ in the discrete-time 
readiness model are assigned an AR(1) prior on the original scale, 
denoted $\AR(\eta_\phi, \rho_\phi, \sigma_\phi)$, 
with analogous priors on $(\eta_\phi, \rho_\phi, \sigma_\phi)$.

\section{Supplementary Simulation Details}
\label{supp:simulation-details}

\subsection{Data-generating process}

This section describes the data-generating process~(DGP) used in the simulation study.
For each replication, we simulate $n$ independent patients and 
follow them through a sequence of post-surgery encounters until 
death or administrative censoring.

\subsubsection{Baseline and first post-surgery encounter}

For patient $i$, we generate baseline covariates 
$\bsL_{i0} = (L_{i01}, L_{i02}, L_{i03})$ with
$L_{i01}, L_{i02} \sim \Normal(0, 1)$ and $L_{i03} \sim \Bern(0.5)$.
We then generate the first gap time and event type from a 
cause-specific Weibull model,
\begin{equation*}
    h(w_{i1}, \delta_1 = k \mid \bsL_{i0}) = h_{01k}(w_{i1}) \exp(\bsL_{i0}' \bsvarphi_{1k}), \quad
    h_{01k}(t) = \lambda_{1k} \nu_{1k} t^{\nu_{1k} - 1}, \quad
    k \in \qty{1, 2},
\end{equation*}
together with an independent censoring time 
$W_{C_{i1}} \sim \Exp(\lambda_C)$.
We generate time-varying covariates at the encounter via
\begin{equation*}
    L_{i11} \sim \Normal(\bsL_{i0}' \bsomega_{L_{11}}, \sigma_{L_{11}}^2), \quad
    \logit\qty{P(L_{i12} = 1 \mid \bsL_{i0})} = \bsL_{i0}' \bsomega_{L_{12}}.
\end{equation*}
Treatment is assigned according to
\begin{equation*}
    \logit\qty{P(A_{i1} = 1 \mid \bsH_{i1})} = \bsH_{i1}' \bsxi_1, \quad
    \bsH_{i1} = (\bsL_{i1}, \bsL_{i0}),
\end{equation*}
and cost is generated from a gamma model with mean
\begin{equation*}
    \mu_{i1} = m_{10}(w_{i1}) \exp(A_{i1} \bsbeta_{A_1} + \bsH_{i1}' \bsbeta_{H_1}), \quad 
    Y_{i1} \sim \GammaDist(\mu_{i1}^2 / \zeta_1, \mu_{i1} / \zeta_1),
\end{equation*}
where the baseline mean-cost function is Gompertz: 
$m_{10}(t) = \eta_1 \exp(\alpha_1 t)$.

\subsubsection{Subsequent encounters}

For $j > 1$, conditional on the most recent state 
$\bsS_{ij} = (A_{i(j-1)}, \bsL_{i(j-1)})$, we generate
\begin{equation*}
    h(w_{ij}, \delta_j = k \mid \bsS_{ij}) = h_{0k}(w_{ij}) \exp(\bsS_{ij}' \bsvarphi_k), \quad
    h_{0k}(t) = \lambda_k \nu_k t^{\nu_k - 1}, \quad
    k \in \qty{1, 2},
\end{equation*}
with an independent censoring time 
$W_{C_{ij}} \sim \Exp(\lambda_C)$.
Time-varying covariates are generated via
\begin{equation*}
    L_{ij1} \sim \Normal(\bsS_{ij}' \bsomega_{L_1}, \sigma_{L_1}^2), \quad
    \logit\qty{P(L_{ij2} = 1 \mid \bsS_{ij})} = \bsS_{ij}' \bsomega_{L_2}.
\end{equation*}
Treatment assignment follows
\begin{equation*}
    \logit\qty{P(A_{ij} = 1 \mid \bsH_{ij})} = \bsH_{ij}' \bsxi, \quad
    \bsH_{ij} = (\bsL_{ij}, \bsS_{i(j-1)}),
\end{equation*}
and costs are generated from
\begin{equation*}
    \mu_{ij} = m_0(w_{ij}) \exp(A_{ij} \bsbeta_A + \bsH_{ij}' \bsbeta_H), \quad
    Y_{ij} \sim \GammaDist(\mu_{ij}^2 / \zeta, \mu_{ij} / \zeta),
\end{equation*}
with Gompertz baseline mean cost $m_0(t) = \eta \exp(\alpha t)$.

\subsection{Parameter settings and study design}

We omit the treatment readiness indicator $Z_{ij}$ by 
assuming treatment can be initiated at any post-surgery encounter; 
including $Z_{ij}$ increases model complexity and Monte Carlo error by 
introducing additional nuisance parameters.
Administrative censoring is controlled by $\lambda_C$, 
set to $0.1$ or $0.5$ for approximately 10\% or 50\% censoring, respectively.
For each censoring level, we generate 1,000 independent datasets with $n = 1,000$ patients.

For $j \geq 1$, we set the following gap time, covariate, treatment, and cost model parameters:
\begin{gather*}
    (\lambda_{11}, \lambda_{12}, \nu_{11}, \nu_{12}) = (1.3, 0.4, 3.0, 2.3), \\
    \bsvarphi_{11} = (0.7, 0.25, -0.1), \quad
    \bsvarphi_{12} = (-0.2, 0.1, 0.35), \\
    (\lambda_1, \lambda_2, \nu_1, \nu_2) = (1.2, 0.8, 2.8, 2.3), \\
    \bsvarphi_1 = (-0.1, 0.15, 0.45, -0.2, 0.1, 0.35), \quad
    \bsvarphi_2 = (-0.1, 0.25, -0.45, 0.7, 0.25, -0.1), \\
    \bsomega_{L_{11}} = (-0.35, 0.1, 0.1), \quad
    \bsomega_{L_1} = (-0.15, 0.2, 0.1, -0.35, 0.1, 0.1), \\
    \sigma_{L_{11}} = \sigma_{L_1} = 1.0, \\
    \bsomega_{L_{12}} = (0.15, -0.25, 0.4), \quad
    \bsomega_{L_2} = (-0.2, 0.1, 0.1, 0.15, -0.25, 0.4), \\
    \bsxi_1 = (0.15, -0.6, 0.2, 0.15, -0.4), \quad
    \bsxi = (0.1, 0.25, 0.15, -0.6, 0.35, 0.2, 0.15, -0.4), \\
    (\eta_1, \alpha_1) = (\eta, \alpha) = (0.8, 0.12), \\
    \bsbeta_{A_1} = \bsbeta_A = 0.15, \\
    \bsbeta_{H_1} = (0.1, 0.45, -0.15, 0.30, -0.05), \quad
    \bsbeta_H = (0.05, 0.15, 0.1, 0.45, 0.1, -0.15, 0.30, -0.05), \\
    \zeta_1 = \zeta = 0.4.
\end{gather*}

\subsection{Estimators compared}

Correctly specified parametric models match the DGP, 
while misspecification is introduced by 
assuming exponential baseline hazards and baseline mean costs.
The discrete-time maximum likelihood comparator follows 
\citet{Spieker2018AnalyzingMedical,Spieker2020NestedComputation}, 
extended to gamma cost outcomes by modeling the baseline mean cost as 
piecewise constant over discrete intervals.
All Bayesian procedures are implemented in \texttt{Stan}~\citep{Carpenter2017StanProbabilistic}; 
frequentist comparators use maximum likelihood estimation.

\section{Supplementary Application Details}
\label{supp:application-details}

\begin{figure}[tbp]
    \centering
    \includegraphics[width=0.45\textwidth]{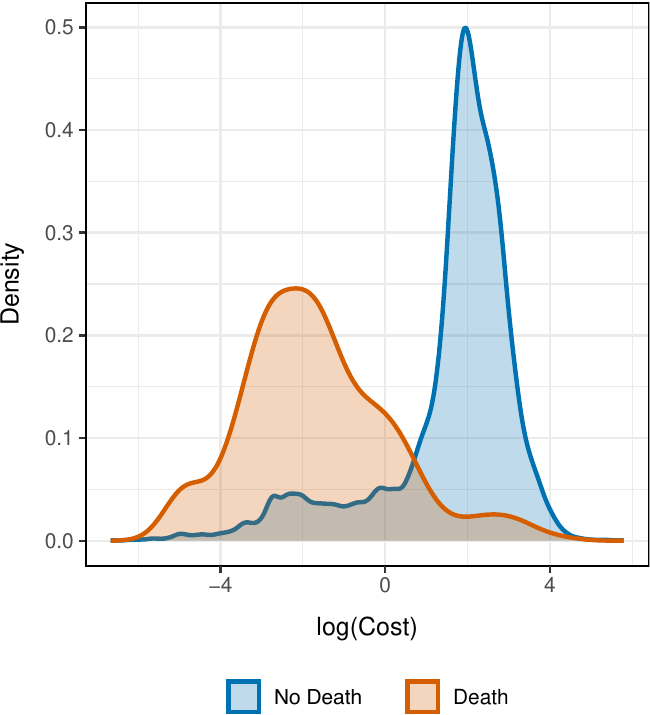}
    \caption{Kernel density estimates of log-transformed encounter costs, $\log Y_j$, stratified by event type: non-death encounters $\delta_j \in \qty{0, 1}$ versus death encounters $\delta_j = 2$.}
    \label{fig:event-specific-cost-density}
\end{figure}

Figure~\ref{fig:event-specific-cost-density} shows kernel density estimates 
of log-transformed encounter costs, stratified by event type.
On the log scale, costs recorded at regular encounters 
(including administratively censored encounters) tend to exceed those observed at death.
To accommodate these systematic differences, the main text includes an 
event-specific multiplicative shift parameter in the cost model.

\begin{figure}[tbp]
    \centering
    \includegraphics[width=0.6\textwidth]{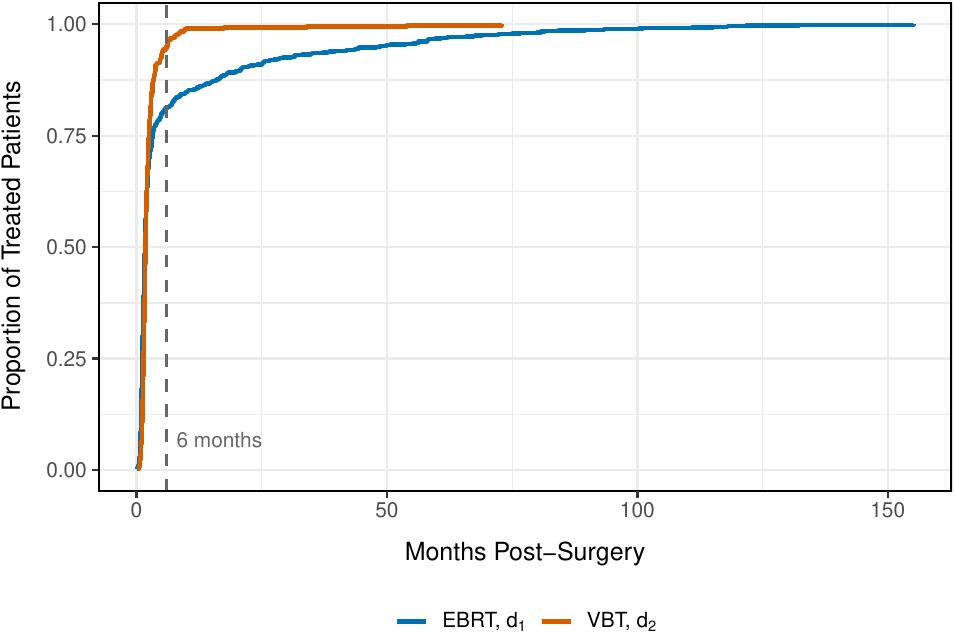}
    \caption{Kaplan-Meier curves for time to adjuvant radiation initiation within treatment groups.}
    \label{fig:treatment-positivity}
\end{figure}

Figure~\ref{fig:treatment-positivity} shows Kaplan-Meier curves for 
time to adjuvant radiation initiation, stratified by treatment group.
Across groups, at least 75\% of patients initiate treatment within six months, 
providing empirical support for the treatment-positivity condition used for identification.

Our application models closely follow the simulation setup, with two modifications.
First, models condition on the observed history used in the main analysis, including the most recent gap time $W_j$, covariates $\bsL_j$, treatment $A_j$, and prior cost information (as available through $\bsS_j$ and $\bsH_j$).
Second, because a fraction of patients have zero recorded cost at their terminal (death) encounter, we use a hurdle component for at-death costs.
Specifically, we model the probability of a nonzero terminal cost using a Bernoulli regression, and, conditional on being positive, model the cost magnitude with the same likelihood used elsewhere (that is, gamma).
Both parts of the hurdle model adjust for the same set of confounders.

\end{document}